\documentclass[aps,twocolumn,preprintnumbers,amsmath,amssymb,prd]{revtex4}
\usepackage{natbib}
\usepackage{amsmath}
\usepackage{amssymb}
\usepackage{graphicx}
\usepackage{bm}

\def\sss{\scriptscriptstyle}
\def\ts{\textstyle}

\def\be{\begin{equation}}
\def\ee{\end{equation}}
\def\ba{\begin{eqnarray}}
\def\ea{\end{eqnarray}}
\def\Mpc{h^{-1}\,{\rm Mpc}}
\def\kpc{h^{-1}\,{\rm kpc}}
\def\Gpc{h^{-1}\,{\rm Gpc}}
\def\Msol{h^{-1}\,{\rm M_{\odot}}}
\newcommand{\mx}{\mbox}
\newcommand{\bom}{\boldmath}

\def\apjl{Astrophys.\ J.\ Lett.}  \def\mnras{Mon.\ Not.\ R.\ Astron.\ Soc.}  

  \def\aap{Astron.\ Astrophys.}
\def\apj{Astrophys.\ J.}  \def\aj{Astron.\ J.}  \def\apjs{Astrophys.\ J. Supp.}  
  \def\prd{Phys.\ Rev.\ D}
  \def\physrep{Phys. Rep.}

\begin{document}

\title{The impact of correlated projections on weak lensing cluster counts}

\author{Laura Marian$^{1}$, Robert E. Smith$^{2}$, Gary M. Bernstein$^{3}$\vspace{0.2cm}}
\affiliation{
$^{1}$Argelander-Institut f\"{u}r Astronomie, Universit\"{a}t Bonn, Bonn, D-53121, Germany\\
$^{2}$Institute for Theoretical Physics, University of Z\"{u}rich, Z\"{u}rich, CH 8037, Switzerland\\  
$^{3}$Department of Physics and Astronomy, University of Pennsylvania, Philadelphia, PA 19104, USA}


\begin{abstract}
Large-scale structure projections are an obstacle in converting the
shear signal of clusters detected in weak-lensing maps into virial
masses. However, this step is not necessary for constraining cosmology
with the shear-peak abundance, if we are able to predict its
amplitude. We generate a large ensemble of $N$-body simulations
spanning four cosmological models, with total volume $V_{\rm
  tot}\sim1[{\Gpc}]^3$ per model. Variations to the matter density
parameter and amplitude of fluctuations are considered. We measure the
abundance of peaks in the mass density projected in $\approx$100
$\Mpc$ slabs to determine the impact of structures spatially
correlated with the simulation clusters, identified by the 3D
friends-of-friends algorithm. The halo model shows that the choice of
the smoothing filter for the density field is important in reducing
the contribution of correlated projections to individual halo
masses. Such contributions are less than 2\% in the case of the
optimal, compensated filter used throughout this analysis. We measure
the change in the mass of peaks when projected in slabs of various
thicknesses. Peaks in slabs of 26 $\Mpc$ and 102 $\Mpc$ suffer an
average mass change of less than 2\% compared to their mass in slabs
of $51\,\Mpc$. We then explore the cosmology dependence of the
projected-peak mass function, and find that, for a wide range of slab
thicknesses $(<500\,\Mpc)$, it scales with cosmology in exactly the
same way as the 3D friends-of-friends mass function and the
Sheth-Tormen formula. This extends the earlier result of
\citet{2009ApJ...698L..33M}. Finally, we show that for all
cosmological models considered, the low and intermediate mass bins of
the peak abundance can be described using a modified Sheth-Tormen
functional form to within 10\%-20\% accuracy.
\end{abstract}
\maketitle
\section{Introduction}\label{Intro}
For more than a decade, weak gravitational lensing (WL) has been
considered a powerful probe for testing the cosmological framework.
The basic idea of gravitational lensing is that on its way to the
observer, the light from distant sources is deflected by the
intervening matter along the line of sight. On average these
deflections are very small, so the result in most cases is that a
circular source galaxy looks slightly elliptical, i.e. the image of
the source is sheared. Images can also be magnified. For a
quantitative idea, the shear is typically of about 2\%, hence the name
of WL. By measuring the shear, we can map the 3D matter distribution
of the universe in an unbiased way, independent of baryonic matter
tracers. Indeed, the largest surveys to date aimed at measuring cosmic
shear, from CTIO (\citet{2003AJ....125.1014J, 2006ApJ...644...71J})
and more recently the CFHTLS (\citet{2006ApJ...647..116H,
  2006A&A...452...51S}), have demonstrated the potential of WL to
place important constraints on the cosmological model. In particular,
the CTIO results were among the first to indicate that the
normalization of mass fluctuations was lower ($\sigma_8\approx0.8$)
than the WMAP1 (\textcite{2003ApJS..148..175S}) analysis suggested
($\sigma_8\approx0.9$), a finding confirmed now by the WMAP5 analysis
(\citet{2009ApJS..180..330K}).

\par In this study, we concentrate on WL-detected clusters. WL
identifies clusters as peaks in shear maps. The cluster abundance is
one of the four most promising tools to measure dark energy
(\citet{2006astro.ph..9591A}), together with supernova surveys,
baryonic acoustic oscillations, and WL surveys. The cluster abundance
is sensitive to the redshift dependence of the angular diameter
distance, the time dependence of the expansion rate and the growth
rate of structure. There are three very appealing features of
WL-detected clusters:
\begin{enumerate}
\item The shear is straightforward to predict, given a model of
  structure formation. This is not the case for X-ray,
  Sunyaev-Zeldovich, or optical clusters, for which the conversion
  from measurable quantities---flux, temperature, decrement---to
  virial masses depends on the complex gas physics of baryons in
  clusters. Cosmological constraints are likely to be biased if the
  mass-observable relation evolves with redshift in a way that mimics
  cosmological changes.
\item An immediate consequence of the first point is that shear peaks
  are easily studied from numerical simulations of dark matter, since
  they are not very sensitive to the baryonic matter
  distribution. This is a huge advantage over the other cluster
  methods, because dark matter simulations are relatively cheap to
  produce with the current computer technology, whereas simulating the
  evolution of baryons depends on poorly known astrophysics and is
  computationally very demanding and expensive.
\item Finally, shear peaks come ``for free'', i.e. a survey designed
  to measure the ``cosmic shear'' 2-point function can detect clusters
  without much additional observational effort.
\end{enumerate} 

\par One issue that has raised concerns about the utility of WL
clusters is the projection effect: the 2D shear maps encode the 3D
matter distribution information. If a cluster produces a detectable
signal in a lensing map, then to this signal contributes not only the
mass of the cluster, but also the matter along the line of sight. This
matter, external to the cluster itself, can be either nearby it, and
therefore spatially correlated with it, or it can be at such a large
separation that there is no correlation with the cluster. Hence, we
can divide contributions from external matter along the line of sight
into \mx{\it{correlated}} and \mx{\it{uncorrelated
    projections}}. Thus, reconstructing the mass of a cluster can be
challenging as there is no predefined way to separate its shear signal
from that induced by other structures on the same line of sight.

\par Many theoretical and numerical endeavors have been aimed at
estimating the magnitude of this effect, its impact on cluster mass
reconstructions, and on the cosmological constraints derived from the
WL-detected cluster abundance. Here we just mention a few of these
works. \citet{2001A&A...370..743H} used the aperture mass statistic,
introduced by \citet{1996MNRAS.283..837S}, to evaluate the uncertainty
in cluster mass estimates due to the uncorrelated large-scale
structure (LSS) contamination. For large clusters at $z=0.1$, they
found the uncertainty to be \mx{$\approx$} 6\%.

\par Other numerical studies have focused on the efficiency of WL
surveys to detect clusters. \citet{2004MNRAS.350..893H} compared the
signal-to-noise (S/N) of the peaks measured from simulated convergence
maps to the Navarro-Frenk-White (NFW) (\citet{1997ApJ...490..493N})
S/N of the 3D clusters producing the peaks. They found scatter and
bias in the predicted vs. measured S/N relation, which they explained
through both departures of the 3D halo density profiles from the
spherically symmetric NFW, as well as LSS
projections. \citet{2005ApJ...635...60T} also attributed a similar
scatter to the asphericity of halos, and to halo
substructure. \citet{2005NewA...10..676D} concluded that individual
cluster shear measurements are contaminated by projections of very
small clusters along the line of sight, i.e. halos with masses smaller
than $10^{12}\,\Msol$. \citet{2005ApJ...624...59H} performed an
extensive numerical study of the WL-cluster detection. The authors
used tomographic information to determine the redshifts of the lenses,
along with an optimal filter to enhance the cluster detection. One of
their results was that even in the absence of the intrinsic
ellipticity noise -- which is known to produce spurious peaks, see for
instance \citet{2004MNRAS.350..893H} -- about 15\% of the most
significant peaks were due to LSS projections, and not to any halo in
particular. \citet{2004PhRvD..70b3008D} and
\citet{2005A&A...442..851M} have independently proposed optimal linear
estimators with a shape designed to filter out the LSS
contributions. \citet{2005A&A...442..851M} used $N$-body simulations
to compare their estimator, which is formally an aperture mass
estimator, to the standard aperture mass filter, proposed by
\citet{1998MNRAS.296..873S} for measurements of cosmic shear. The
numerical tests have indicated their filter to be more efficient than
the standard aperture mass at discarding 'false positives', i.e. those
shear peaks arising from LSS projections. \citet{2007A&A...462..875S}
and \citet{2007A&A...462..473M} detected clusters in shear data and
found the performance of the optimal filter of
\citet{2005A&A...442..851M} to be similar to that of the aperture mass
statistic used with a NFW-matched filter. Another example of cluster
detection in shear data is the work of \citet{2009arXiv0904.2185A}.

\par \citet{2006PhRvD..73l3525M} also used an optimal shear filter to
forecast WL-cluster constraints for future ground- and space-based
surveys. They found that cosmological constraints are not severely
altered by uncorrelated LSS projections. One of the critical
assumptions that forecasts adopt (see also
\citet{2004PhRvD..70l3008W}) is to consider the Sheth-Tormen (ST)
(\citet{1999MNRAS.308..119S}) mass function a proxy for the shear-peak
mass function. But is the shear-peak mass function as sensitive to
cosmology as the 3D one? In a pioneering study by
\citet{1999A&A...351..815R}, reasonable agreement was found between
the analytic predictions from the model of \citet{1999MNRAS.302..821K}
and peak counts measured from numerical simulations. However, the
small sample volumes studied, $V_{\rm tot}\sim 10^{-3}[{\Gpc}]^{3}$,
limited the statistical power of the conclusions, especially for
application to future WL data sets. In a recent analysis
(\citet{2009ApJ...698L..33M}), we used a very large ensemble of
numerical simulations to pursue the issue further. We showed that the
2D mass function of peaks projected in thin slabs scales with
cosmology in the same way as the measured 3D mass function, and also
as the ST prediction. Owing to the large total volume $V_{\rm tot}\sim
1[\Gpc]^3$ per cosmological model, the result was established at very
high statistical significance. This suggests that for the purpose of
constraining cosmology, one can avoid the difficult task of converting
shear peaks into cluster masses. If the cosmology dependence of the
shear peaks is understood, one needs only measure their abundance to
obtain the desired constraints. \citet{2009arXiv0906.3512D} took this
approach numerically and derived constraints from shear peaks measured
from simulations with the aperture mass statistic. See also the
related work of \citet{2009arXiv0907.0486K}, and
\citet{2009arXiv0904.2995P}.

In this paper we continue our investigation of the effect of
correlated mass structures on the projected-peak mass function and its
power to constrain cosmological models. 

To achieve this goal, we use numerical simulations. For each
simulation box we divide the matter density field across the line of
sight up into slabs. Our observable is the density field projected
into such slabs. If the slabs are thin, this observable is equivalent
to the WL convergence -- defined as the surface mass density scaled by
the critical surface mass density, $\Sigma_{\rm crit}$. Even when the
slabs get thicker (256, 512 $\Mpc$), the critical surface mass density
varies only a few percents of its central value for typical lenses and
sources in the $\Lambda$CDM model, as pointed out by
\citet{2001ApJ...547..560M}. The reason we are interested in slabs is
that, by definition, we expect only the matter in a limited spatial
volume around a main cluster lens to give rise to correlated
projections. We underline that uncorrelated projections are the
subject of a future study, and not of this study.

We extend the analysis of \citet{2009ApJ...698L..33M} in a number of
important ways. Firstly, we apply the halo model of structure
formation to quantify the expected mass increase arising from
correlated structures along the line of sight. Secondly, we generalize
our earlier results through studying the projected peak mass function
in slabs of various thicknesses. Thirdly, we show how the
projected-peak mass function can be modeled. 


\par The paper is structured as follows. In section \S\ref{Sims} we
present the numerical simulations that our study is based upon; in
\S\ref{HMT} we use the halo model to predict the change in peak masses
due to correlated projections and in \S\ref{Filter} we describe the
optimal filter that we apply to the data. In \S\ref{Algorithm} we show
how we find and select the density peaks. Section \S\ref{Results}
contains the main results of our study, and in \S\ref{Conclusions} we
draw our conclusions. Finally, we would like to clarify the mass
definitions that we use throughout this work. We shall refer to those
masses measured using full 3D information, as 3D masses ($M_{\rm
  3D}$). Similarly, 2D masses are measured from 2D data ($M_{\rm
  2D}$). Depending on the context, $M_{\rm 3D}$ is usually either the
virial mass defined by ST, or the mass defined by the
friends-of-friends (FoF) algorithm of \citet{1985ApJ...292..371D}. We
shall specify the meaning of the notation $M_{\rm 3D}$ in the context
where we use it. $M_{\rm 2D}$ will always mean the mass of a projected
peak, and the way we assign masses to projected peaks is described in
section \S\ref{Filter}.
\section{Numerical simulations}\label{Sims}
We have generated an ensemble of 32 $N$-body simulations for 4
different cosmological models, with 8 realizations per model. The
simulations were run using {\tt Gadget-2}
(\citet{2005MNRAS.364.1105S}), and they have $400^{3}$ particles of
mass $m_{\rm p}\approx\, 10^{11}\,\Msol$ in a simulated volume of
$512^{3} \,(\Mpc)^{3}$. The full details are in Table
~\ref{table:sims}. There are 18 snapshots, logarithmically spaced in
the scale factor $a$, between redshifts 50 and 0. We used the {\tt
  2LPT} code of \citet{1998MNRAS.299.1097S} to generate the initial
conditions for the density fluctuations. For each of the 8
realizations, the initial conditions for the 4 cosmologies are the
same. Thus, the cosmic variance on the comparison of measured
quantities in different cosmologies will be minimal. The transfer
function was taken from {\tt cmbfast} (\citet{1996ApJ...469..437S})
(the output at redshift 0). We identified the simulation halos using
the FoF algorithm, with $b=0.2$, where $b$ is the fraction of the
inter-particle separation. The smallest of our halos have 30
particles. Our fiducial cosmology is described by Model 1 in
Table~\ref{table:sims}. We shall refer to the other models as the
variational cosmologies.


\begin{table}
\caption{Numerical simulation details. Top-table columns are: matter,
  dark energy, and baryon density parameters; normalization of
  fluctuations; mass per particle in units of $(10^{11}\,\Msol)$;
  number of realizations. Bottom-table columns show: box length; dark
  energy equation of state $w=p_{\rm \sss DE}/\rho_{\rm \sss DE}$; number of particles;
  Hubble's constant in units of 100 ${\rm km\,s^{-1}\,Mpc^{-1}}$; baryon
  fraction.}
\label{table:sims}
\vspace{0.2cm}
\begin{tabular}{c|cccccc}
        & $\Omega_{\rm m}$ \hspace{0.2cm}& $\Omega_{\rm DE}$ \hspace{0.2cm} & $\Omega_{\rm b}$ \hspace{0.2cm}& $\sigma_{\rm 8}$  & $m_{\rm p}$ & $N_{\rm run}$ \\
\hline
Model 1 \hspace{0.1cm}&   \hspace{0.1cm} 0.27 \hspace{0.2cm} &  0.73 \hspace{0.2cm} &  0.0460 \hspace{0.2cm}  & 0.9 & 1.57  \hspace{0.2cm} & 8 \hspace{0.2cm} \\
Model 2  \hspace{0.1cm}&   \hspace{0.1cm} 0.22 \hspace{0.2cm} &  0.78 \hspace{0.2cm} &  0.0375 \hspace{0.2cm} & 0.9  & 1.28  \hspace{0.2cm} & 8 \hspace{0.2cm} \\
Model 3  \hspace{0.1cm}&   \hspace{0.1cm} 0.27 \hspace{0.2cm} &  0.73 \hspace{0.2cm} &  0.0460 \hspace{0.2cm} & 0.75  & 1.57 \hspace{0.2cm} & 8 \hspace{0.2cm} \\
Model 4  \hspace{0.1cm}&   \hspace{0.1cm} 0.32 \hspace{0.2cm} &  0.68 \hspace{0.2cm} &  0.0545 \hspace{0.2cm}  & 0.9  & 1.86  \hspace{0.2cm} & 8  \hspace{0.2cm} \\
\hline
\end{tabular}
\begin{tabular}{ccccc}
$L=512 \,\Mpc$\hspace{0.1cm} & $w=-1$ & $N_{\rm p}=400^{3}$&
  $h=0.72$ & $f_{\rm b}$=17\% \hspace{1cm}\\ \hline
\end{tabular}
\end{table}
\section{Theoretical prediction of the correlated projection effect} 
\label{HMT}
In this section we develop a theoretical prediction for the correlated
projection effect. We base our calculation for the correlated
projections on the halo model, and we rigorously test it against
numerical simulations. Our approach is the following: we test the
strength of the projection effect in slabs of various thickness. For a
cluster at the center of such a slab, we expect that the contribution
from structures correlated with it will flatten out as we increase the
slab thickness. Only those objects within the correlation length of
the cluster ($r_0 \sim 30\,\Mpc$ for clusters more massive than
$10^{14}\,\Msol$) will affect its lensing signal. As the slab thickness
becomes large -- a few hundreds Mpc -- it is reasonable to assume that
chance or uncorrelated projections will start to contribute.

If the large-scale distribution of the mass in the Universe were
perfectly described by the halo model, then all the mass would be
contained in halos with universal properties: spherical symmetry, a
universal density profile, and a mass function; see
\citep{2002PhR...372....1C} for a halo model review. In this
framework, we would like to estimate the average contribution to the
projected density of a halo of mass $M_{1}$ from structures correlated
with it.  The probability of finding a halo with mass in the interval
$(M_{2}, M_{2}+ dM_{2})$, within a range $(\mx{\bom{$r$}},
\mx{\bom{$r$}}+ d^{3}r)$ of the center of $M_{1}$ is given by:
\be
\mathcal P(\mx{\bom $r$,}M_{2} | M_{1}) dM_{2}\,d^{3}r=
n(M_{2})\,[1+\xi_{\times}(\mx{\bom $r$}, M_{2}| M_{1})]dM_{2} \,d^{3}r,   
\label{eq:probM2}
\ee
where $n(M_{2})$ is the halo mass function, i.e. the comoving number
density of halos of mass $M_{2}$ per unit mass, and $\xi_{\times}$ is
the cross-correlation between the center of halo $M_{1}$ and the mass
distribution of $M_{2}$. The expected density profile of matter
\mx{\it{external}} to the halo $M_{1}$ is:
\be \langle \rho_{\rm ext}(\mx{\bom $r$}|M_{1})\rangle= \int dM_{2}\,M_{2}\,
n(M_{2})\,[1+\xi_{\times}(\mx{\bom $r$}, M_{2}|M_{1})].
\label{eq:3dAve}
\ee
The cross-correlation function can be obtained from the cross-power
spectrum:
\be
\xi_{\times}(\mx{\bom $r$}, M_{2}| M_{1})=
\int\frac{dk}{2\pi^{2}}\,k^{2}P_{\times}(k, M_{2}| M_{1})\, j_{0}(kr),
\ee
where $j_{0}(x)=\sin(x)/x$ is the zero-order spherical Bessel
function. The halo center cross-power spectrum, at leading order in
the halo perturbation theory, is:
\be
P_{\times}(k, M_{2}| M_{1})=b(M_{1})\,b(M_{2})\,U(k, M_{2}) P(k)\ ,
\label{crossPower}
\ee
where $b(M)$ is the first order linear bias parameter, $P(k)$ is the
nonlinear matter power spectrum. $U(k,M)$ is the Fourier transform of
the 3D density profile of a halo with mass $M$:
\ba
U(k, M) & \equiv & \frac{\rho(k, M)}{M}, \hspace{3.13cm} \nonumber  \\
& = & \frac{4\pi}{M}\int_{0}^{R_{\rm vir}} dr \,r^{2} \rho(r, M) \,j_{0}(kr)\ .
\label{eq:U}
\ea
\begin{figure}
\centering \includegraphics[scale=0.4]{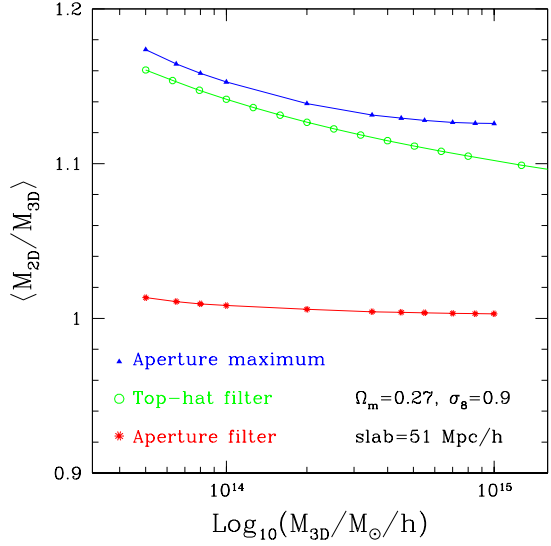}
\caption {The change in mass due to correlated projections as a
  function of the Sheth-Tormen virial mass. The starred red symbols
  correspond to the optimal compensated filter described in
  section~\S\ref{Filter}, and the empty green circles to the top-hat
  filter. These curves are for a slab thickness of 51 $\Mpc$. The blue
  triangles show the maximum correlated projection effect for the
  compensated filter, as given by Eq.~(\ref{eq:M_corr_max}). This plot
  corresponds to the fiducial cosmology.}
\label{fig:HM_PRED}
\end{figure}
In the above, we have assumed that the 3D density profile of the halo
is cut off at the virial radius $R_{\rm vir}$.  
The expected projected mass density external to the halo with mass $M_{1}$
is the integral of Eq.(\ref{eq:3dAve}) along the projection axis -- for
example the $\hat z$ direction:
\be
\langle \Sigma_{\rm ext}(\mx{\bom $r_{\perp}$}|M_{1})\rangle=
2\, \int_{\sqrt{R_{\rm vir}(M_{1})^2-r_{\perp}^2}}^{L/2} dz \,\langle \rho_{\rm ext}(z,\mx{\bom $r_{\perp}$}|M_{1})\rangle\,,
\label{eq:2dAve1}
\ee
where $\mx{\bom $r_{\perp}$}$ is the 2D position vector transverse to
the line of sight, and $L$ is the slab thickness.  For convenience, we
have taken the center of halo $M_{1}$ to be at half the slab's
thickness.
Combining the above expressions, we can write the last equation as a
sum of contributions from uniformly-distributed matter, uncorrelated
with the cluster $M_{1}$, and from clustered matter, correlated with
the halo:
\be
\langle M_{\rm ext}(M_{1})\rangle=
\langle M_{\rm uni}(M_{1})\rangle + \langle M_{\rm corr}(M_{1})\rangle.
\label{eq:2dAve2}
\ee
To predict the change in the mass of a projected peak due to both
these terms, we need to convolve Eq.~(\ref{eq:2dAve1}) with a filter,
that we shall generically denote by $W$: 
\ba
\langle M_{\rm uni}(M_{1})\rangle & = & 
2\,\rho_{\rm m}\int d^{2}r_{\perp}W(\mx{\bom $r_{\perp}$})\int_{\sqrt{R_{\rm vir}(M_{1})^2-r_{\perp}^2}}^{L/2} dz\,;
\nonumber \\
\label{eq:M_uncorr}
\\
\langle M_{\rm corr}(M_{1})\rangle & = & 2 b(M_{1})\int dM_{2}\,M_{2} \,b(M_{2})\, n(M_{2})
\nonumber \\
& \times  & 
\int \frac{dk}{2\pi^{2}}\, k^{2} \,U(k, M_{2})\, P(k) \int d^{2} r_{\perp} W(\mx{\bom $r_{\perp}$})
\nonumber \\
& \times &  
\int_{\sqrt{R_{\rm vir}(M_{1})^2-r_{\perp}^2} }^{L/2}\!\! dz\, j_{0}(k\,\sqrt{r_{\perp}^{2}+z^{2}}) \ ,
\label{eq:M_corr}
\ea
where $\rho_{\rm m}$ is the mean matter density of the Universe. $W$
is a dimensionless function, so the right-hand side members of the
above equations have mass dimension.  If $W$ is a compensated filter
(see section \S~\ref{Filter}) then the right-hand side of
Eq.~(\ref{eq:M_uncorr}) is very small and negative; it would be in
fact zero but for the dependence on $r_\perp$ of the lower limit of
the line-of-sight integral. Throughout this analysis we are using a
compensated filter, so we shall drop the contribution of $\langle
M_{\rm uni} \rangle$ since it is tiny compared to $\langle M_{\rm
  corr} \rangle$. Note that an upper bound for the magnitude of the
impact of correlated projections can be obtained if one takes the
lower limit of the line-of-sight integral in Eq.~(\ref{eq:M_corr}) to
be 0, and the slab thickness to be infinite. In that case, only the
modes transverse to the line of sight contribute to $\langle M_{\rm
  corr}\rangle$ and one can rewrite Eq.~(\ref{eq:M_corr}) in the following
way: 
\ba 
\langle M^{\rm max}_{\rm corr}(M_{1})\rangle & = & b(M_{1})\int dM_{2}\,M_{2}
\,b(M_{2})\, n(M_{2}) \nonumber \\ & \times & \int dk_{\perp}\,
k_{\perp}\,U(k_{\perp}, M_{2})\, P(k_{\perp})\nonumber\\
 & \times & \int_{0}^{R_{\rm vir}} dr_{\perp} r_{\perp} W(r_{\perp})
J_{0}(k_{\perp}r_{\perp})\,,
\label{eq:M_corr_max}
\ea
where $J_{0}$ is the zero-order Bessel function. To illustrate the
mass change induced by correlated projections on 3D NFW halos, we have
estimated Eq.~(\ref{eq:M_corr}) in two cases: a top-hat filter, and a
compensated, NFW-shaped filter that we describe in \S\ref{Filter} and
that we used for the entire study. The limits of the $k$-integral were
adjusted to match the volume of our simulations. For the top-hat
filter, Eq.~(\ref{eq:M_corr}) can be written as:
\ba \langle M_{\rm corr}(M_{1})\rangle=2 b(M_{1})\rho_{\rm
  m}\int_{0}^{\infty}dk\,P(k)\,\times\hspace{2cm}\nonumber\\ \times \int_{R_{\rm
    vir}(M_{1})}^{L/2} dz\left[\cos(kz)-\cos(k\sqrt{R_{\rm
      vir}(M_{1})+z^2})\right].
\label{M_corr_th}\hspace{0.3cm}
\ea
Here we have taken $U(k, M_{2})\approx 1$; the integral over $M_2$ was
then evaluated directly, giving $\rho_{\rm m}$. 

Figure~\ref{fig:HM_PRED} shows the change in mass arising from
correlated projections in a 51 $\Mpc$ slab as a function of the virial
mass of halos. The top-hat results are represented with green empty
circles, while the red stars correspond to the compensated filter that
we use throughout this work. The blue triangles are the upper limit
contribution given by Eq.~(\ref{eq:M_corr_max}), and computed with the
compensated filter. The message of the figure is that the filter
choice is crucial in reducing the impact of correlated projections on
the measured peaks: the change in mass is $\approx$ 8 times larger for
the top-hat filter than for the compensated, NFW filter. This is due
to the fact that the latter is very sensitive to cluster-scale modes,
and less so to the larger-scale modes which give rise to projections.
\begin{figure*}
\centering
\includegraphics[scale=0.25]{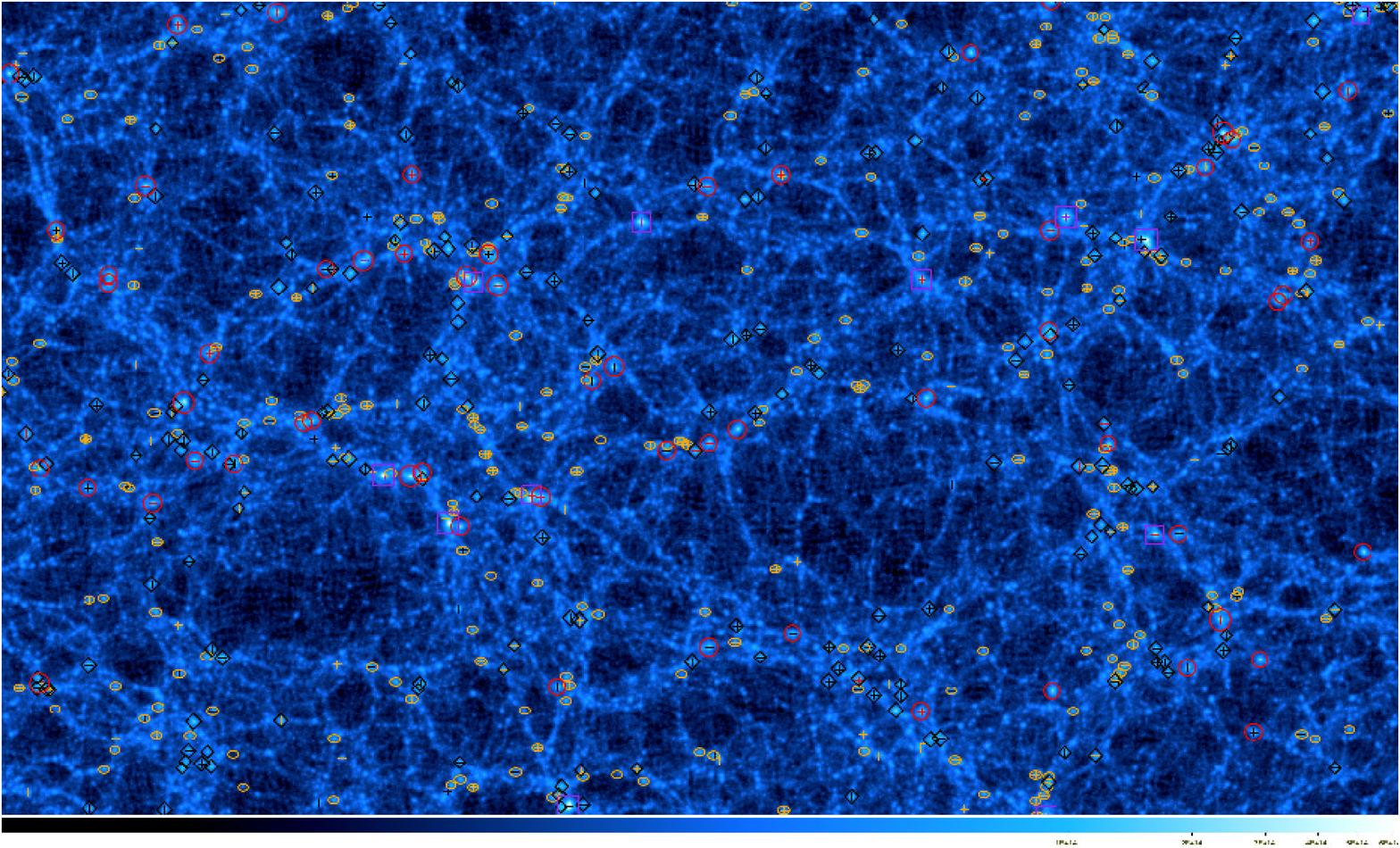}
\caption{The filtered projected density field of a 51 $\Mpc$ slab in
  the fiducial cosmology. The scale is $512 \times 280\,\Mpc$. The
  detected peaks and FoF halos are marked with colored symbols
  depending on their mass range. An enlarged snapshot of this image is
  shown in Figure~\ref{fig:filslabzoom1}.}
\label{fig:filslabzoom0.25}
\end{figure*}
\begin{figure*}
\centering
\includegraphics[scale=0.245]{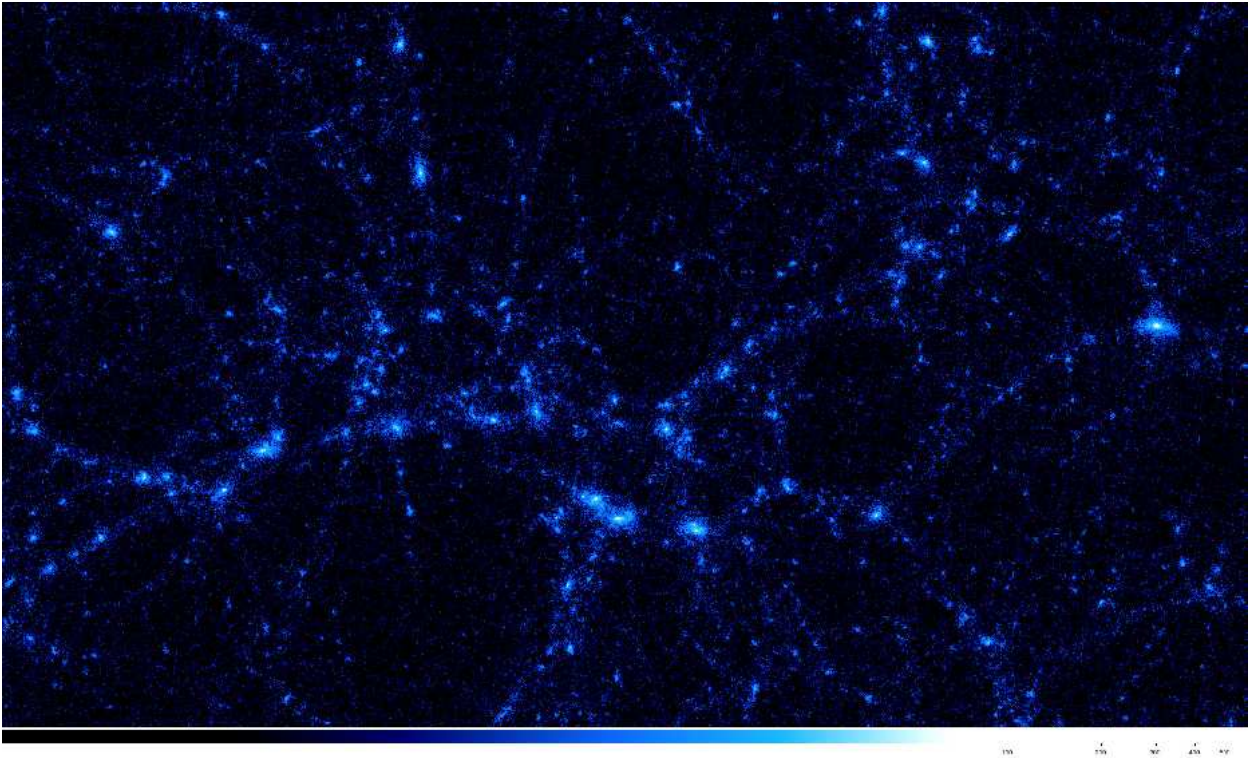}\hspace{1cm}
\caption{The unfiltered projected density of the same 51 $\Mpc$ slab
  shown in Figure~\ref{fig:filslabzoom0.25}. The scale of the picture
  is $130 \times 70\, \Mpc$.}
\label{fig:gridslabzoom1}
\end{figure*}
\begin{figure*}
\centering
\includegraphics[scale=0.245]{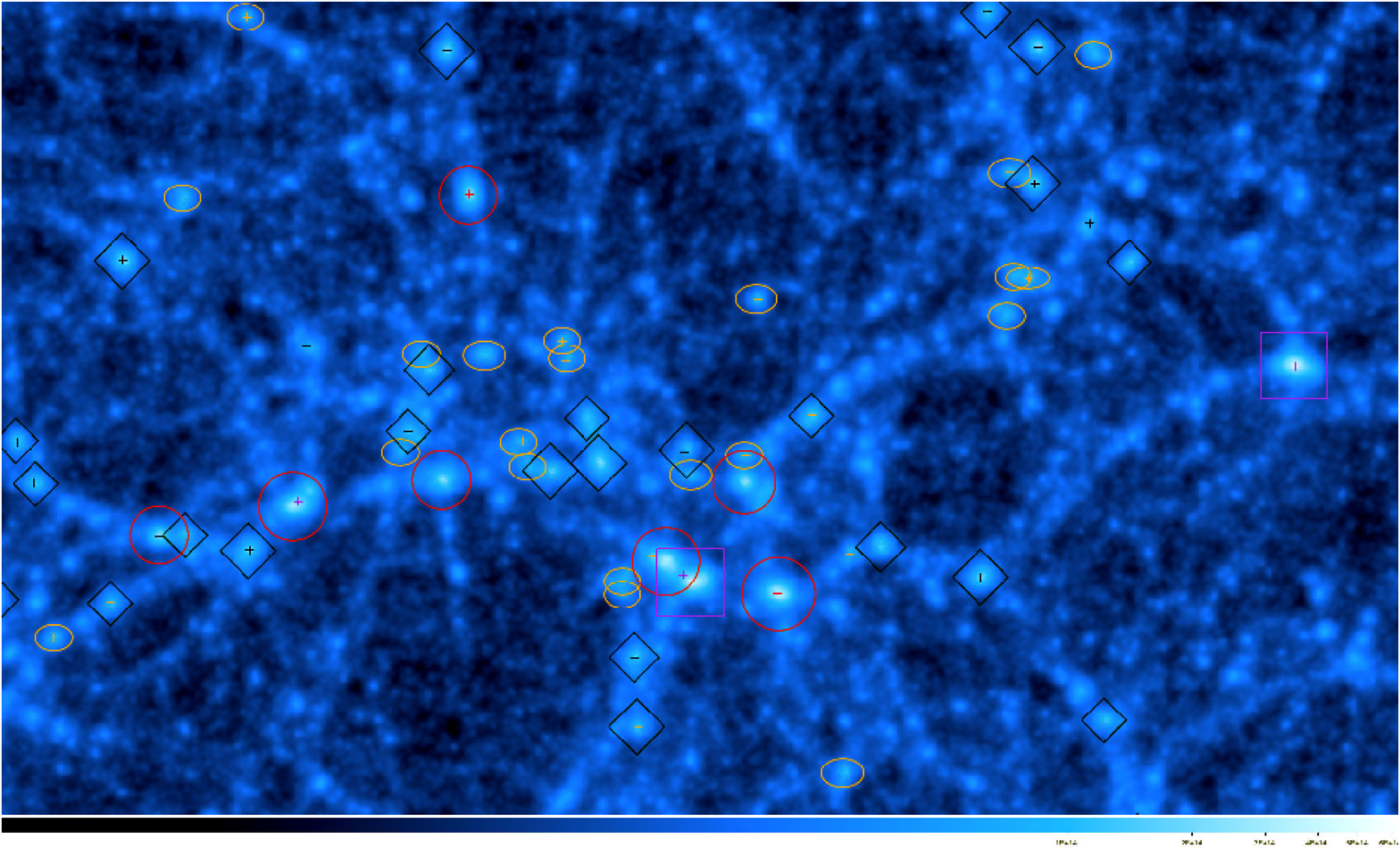}
\caption{Same snapshot as above, but this time the projected density
  field is filtered.  The detected peaks are marked with colored
  symbols depending on their mass range: purple squares for
  $M>5\times10^{14}\,\Msol$; red circles for $5\times10^{14}
 \,\Msol>M>2\times10^{14}\,\Msol$; black rombs for
  $2\times10^{14}\,\Msol>M>8\times10^{13}\,\Msol$; and
  orange ellipses for $8\times10^{13}\,\Msol>M>5\times10^{13}
 \,\Msol$. The FoF halos are represented with crosses and the
  color choice for each mass range is the same as for the peaks. The
  scale of the picture is $130 \times 70\,\Mpc$.}
\label{fig:filslabzoom1}
\end{figure*}
\section{Filtering the projected density field} 
\label{Filter}
In this section we present the filter with which we convolve the
density field, similar to that described in
\citet{2006PhRvD..73l3525M}. We used this filter in section
\S\ref{HMT} to estimate the halo-model prediction of the mass change
induced by correlated projections. Every point of the filtered density
map has an associated mass given by:
\be
M(\mx{\bom $x_{0}$})=\int d^{2}x \, W(\mx{\bom $x_{0}$}-\mx{\bom $x$})
\Sigma(\mx{\bom $x$}),
\label{eq:fil1}
\ee
where $W$ is the filter function and $\Sigma$ denotes the projected
density of dark matter. The filter is compensated
\citep{1996MNRAS.283..837S}, i.e. it vanishes when integrated over the
aperture size. This ensures that no uniform mass sheet contributes to
the measured projected density:
\be 
\int_{0}^{R} d^2x\,W(\mx{\bom $x$})=0, 
\ee 
where $R$ is the aperture radius. The filter is also optimal,
i.e. maximizes the signal-to-noise ($S/N$), and it is tuned to NFW
clusters: an overdensity with the NFW density profile is detected with
a maximum $S/N$. Although the profile of the halos best recovered by
the filter is NFW, the mass is that defined by ST, i.e. an enclosed
overdensity threshold $\rho=\Delta_{\rm vir}\,\rho_{\rm m}$ as opposed
to $\Delta_{\rm vir}\,\rho_{\rm crit}$. Here $\rho_{\rm m}$ is the
mean matter density, $\rho_{\rm crit}$ is the critical density, and
$\rho_{\rm m}=\Omega_{\rm m}\rho_{\rm crit}$. $\Delta_{\rm vir}=200$
for ST and NFW. Thus, in the $\Lambda$CDM model, ST and NFW clusters
have the same density profile, but ST halos have a larger cut-off
radii and concentration parameters than NFW ones. The filter is
normalized to ST virial masses in order to facilitate the comparison
of the 2D peaks with the 3D halos of our simulations, and also with
the ST mass function. There is an agreement of $\approx 10-20\%$
between the FoF and the ST mass functions. As shown in numerous
earlier studies \citep{1996MNRAS.283..837S, 2005A&A...442..851M,
  2006PhRvD..73l3525M}, W is an optimal filter if it is proportional
to the expected measured signal and inverse proportional to the
variance of the noise. If the observable is the convergence field (or
in this case the projected density field), \citet{1996MNRAS.283..837S}
has shown that W is optimal and compensated, if it has the form:
\be
W(\mx{\bom $x$})=\mathcal C_W\left(\Sigma_{\rm \sss ST}(\mx{\bom $x$})-\bar\Sigma_{\rm \sss ST}(R)\right),
\label{eq:fil2}
\ee
with the mean density inside the aperture radius $R$ defined by:
$\bar\Sigma(R)=2/R^2\int_0^R dx\,x\,\Sigma(x)$. We determine the
normalization constant $\mathcal C_W$ from the conditions that the
filter be tuned to the NFW profile and return the ST virial mass:
\be
\mathcal C_W=\frac{M^{\sss \rm ST}_{\rm vir}}{\int d^{2} x \left|\Sigma_{\rm \sss ST}(\mx{\bom $x$})\right|^{2}-\pi R_{\rm vir}^{2}\,\bar\Sigma^2_{\rm \sss ST}(R)}.
\ee
We choose the aperture radius $R$ to be the virial radius associated
with an ST cluster of mass $M^{\sss \rm ST}_{\rm vir}$, i.e. $M^{\sss
  \rm ST}_{\rm vir}=4 \pi \Delta_{\rm vir}\,\rho_{\rm m}\,R_{\rm
  vir}^3/3$. $\Sigma_{\rm \sss ST}$ is the truncated ST projected
density:
\[
\Sigma_{\rm \sss ST}(x)=\ts2\,{r_{\rm s}\,\delta_{\rm c}^{\sss \rm ST}\,\rho_{\rm m}}\,f_{\rm \sss ST}(x).
\]
 $r_{\rm s}$ is the scale radius of the cluster. The characteristic
overdensity of the profile, $\delta_{\rm c}^{\rm \sss ST}$, is related
to the concentration parameter $c_{\rm \sss ST}$ by the condition that
the mean density within $R_{\rm vir}$ should be $\Delta_{\rm vir}
\rho_{\rm m}$:
\[
\delta_{\rm c}^{\sss ST}=\Delta_{\rm vir}\,c_{\sss \rm ST}^{3}/\left( 3\,\left[\log(1+c_{\sss \rm ST})-c_{\rm \sss ST}/(1+c_{\rm \sss ST})\right]\right).
\] 
Finally, $f_{\rm \sss ST}$ is a function which depends on cosmology
only through the concentration parameter, see for instance
\citet{2004MNRAS.350..893H}. Its expression varies depending on
whether it is evaluated at $r<r_{\rm s}$, $r=r_{\rm s}$, $r_{\rm
  s}<r<R_{\rm vir}$, and $r>R_{\rm vir}$:
\be
f_{\rm \sss ST}(\xi)= \left\{ \begin{array}{lcl}
-\frac{\ts{(c_{\rm \sss ST}^{2}-\xi^{2})^{1/2}}}{\ts{(1-\xi^{2})\,(1+c_{\rm \sss ST})}} 
+ \frac{\ts{ \cosh^{-1}\left(\frac{\xi^{2}
+c_{\rm \sss ST}}{\xi(1+c_{\rm \sss ST})}\right)}}{\ts{(1-\xi^{2})^{3/2}}}\,, \vspace{0.2cm}\\
\frac{\ts{(c_{\rm \sss ST}^{2}-1)^{1/2}}}{\ts{3 (1+c_{\rm \sss ST})}}\, 
\left(1+\frac{\ts{1}}{\ts{1+c_{\rm \sss ST}}} \right)\,, \vspace{0.2cm}\\
-\frac{\ts{(c_{\rm \sss ST}^{2}-\xi^{2})^{1/2}}}{\ts{(1-\xi^{2})\,(1+c_{\rm \sss ST})}} 
- \frac{\ts{\cos^{-1}\left(\frac{\xi^{2}
+c_{\rm \sss ST}}{\xi(1+c_{\rm \sss ST})}\right)}}{\ts{(\xi^{2}-1)^{3/2}}}\,, 
 \vspace{0.2cm}\\
0.
\end{array}\right.
\label{eq:STsigma}
\ee
Here $\xi$ is dimensionless, $\xi=r/r_{\rm s}$. The concentration
parameter is a function of the halo redshift and mass; we have
followed the prescription of \citet{2001MNRAS.321..559B} to compute it
and used the method of \citet{2005MNRAS.360..203S} to map $c_{\rm\sss
  NFW}\rightarrow c_{\rm \sss ST}$.
The projected density field smoothed with the filter~(\ref{eq:fil2})
matched to a cluster of mass $M^{\sss \rm ST}_{\rm vir}$ will yield a maximum
signal at the location of an ST cluster center, which we denote as
$M(\mx{\bom $x_{0}$} | M^{\sss \rm ST}_{\rm vir})$. If the cluster's mass were that
of the filter, then the amplitude of the smoothed field at the peak
would be:
\be
M(\mx{\bom $x_{0}$} | M^{\sss \rm ST}_{\rm vir}) = M^{\sss \rm ST}_{\rm vir}
\label{matchmass}.
\ee
In practice, in order to assign a mass to a cluster, we apply several
filters of different masses, and interpolate to find the mass $M^{\sss
  \rm ST}_{\rm vir}$ at which (\ref{matchmass}) would be
satisfied. This is the mass assigned to the peak.
\section{A hierarchical peak finding algorithm}
\label{Algorithm}
As stated earlier, our observable is the projected density field of
dark matter, and our goal is to study the effect of correlated
projections on the 2D-peak mass function. We divide the simulation
cube in several numbers of slabs of equal thickness: 20 (slabs of
approximately 26 $\Mpc$), 10 ($\approx 51\,\Mpc$), 5 ($\approx 102\,
\Mpc$), 2 ($= 256\, \Mpc$), and 1 ($= 512\, \Mpc$). Each set of equal
slabs is analyzed separately, i.e. the projected density field is
filtered, and peaks are identified as the highest maxima in the
filtered map. A selection algorithm is then applied to decide on the
validity and on the final mass of these peaks.

\par One distinguishing feature of our method is that we filter the
density maps recursively, with a hierarchy of matched filters. The
shape of the matched filter depends upon the mass $M^{\sss\rm ST}_{\rm
  vir}$ of the cluster one is attempting to detect. We start with the
highest-mass filter, convolve it with the map, and find all peaks
above the target mass. We repeat the procedure for lower mass
filters. Peaks that have been already found with a higher mass filter
are discarded. Thus, we do not face the problem of `halos-in-halos'
(\citet{2004MNRAS.350..893H}). The last filter applied corresponds to
the detection threshold of $5\times 10^{13}\,\Msol$. Below this
threshold, the mass resolution of the simulations affects the filter
efficiency, as shown in the next section. To the identified peaks we
associate a virial radius given by $R_{\rm vir}=(3M^{\sss\rm ST}_{\rm
  vir}/4\pi \Delta_{\rm vir}\rho_{\rm m})^{1/3}$, as described in the
previous section. If two peaks have distinct centers
(i.e. $\hat{x}-\hat{y}$ coordinates), but overlapping radii of the
same size, we keep both peaks. If the centers are distinct, but one of
the peaks has a smaller radius than the other one, and it is therefore
`enclosed' in the projected disk of the larger peak, we keep the
higher-mass peak and discard the lower-mass one.
Figures~\ref{fig:filslabzoom0.25}, ~\ref{fig:gridslabzoom1}, and
~\ref{fig:filslabzoom1} show filtered and unfiltered maps of one of
the slabs of the fiducial cosmology. Figure~\ref{fig:filslabzoom0.25}
presents a filtered slab, with the detected peaks and FoF halos marked
on the filamentary structure of the projected density
field. Figure~\ref{fig:gridslabzoom1} is a zoom-in of the same slab,
before filtering, and Figure~\ref{fig:filslabzoom1} is a zoom-in after
filtering. In the latter, FoF halos are marked by crosses of different
colors, according to the mass range where they belong, while peaks are
marked by symbols of various shapes and colors -- see the caption for
the mass legend. 
\begin{figure*}
\centering
\begin{minipage}[t]{0.45\linewidth}
\includegraphics[scale=0.4]{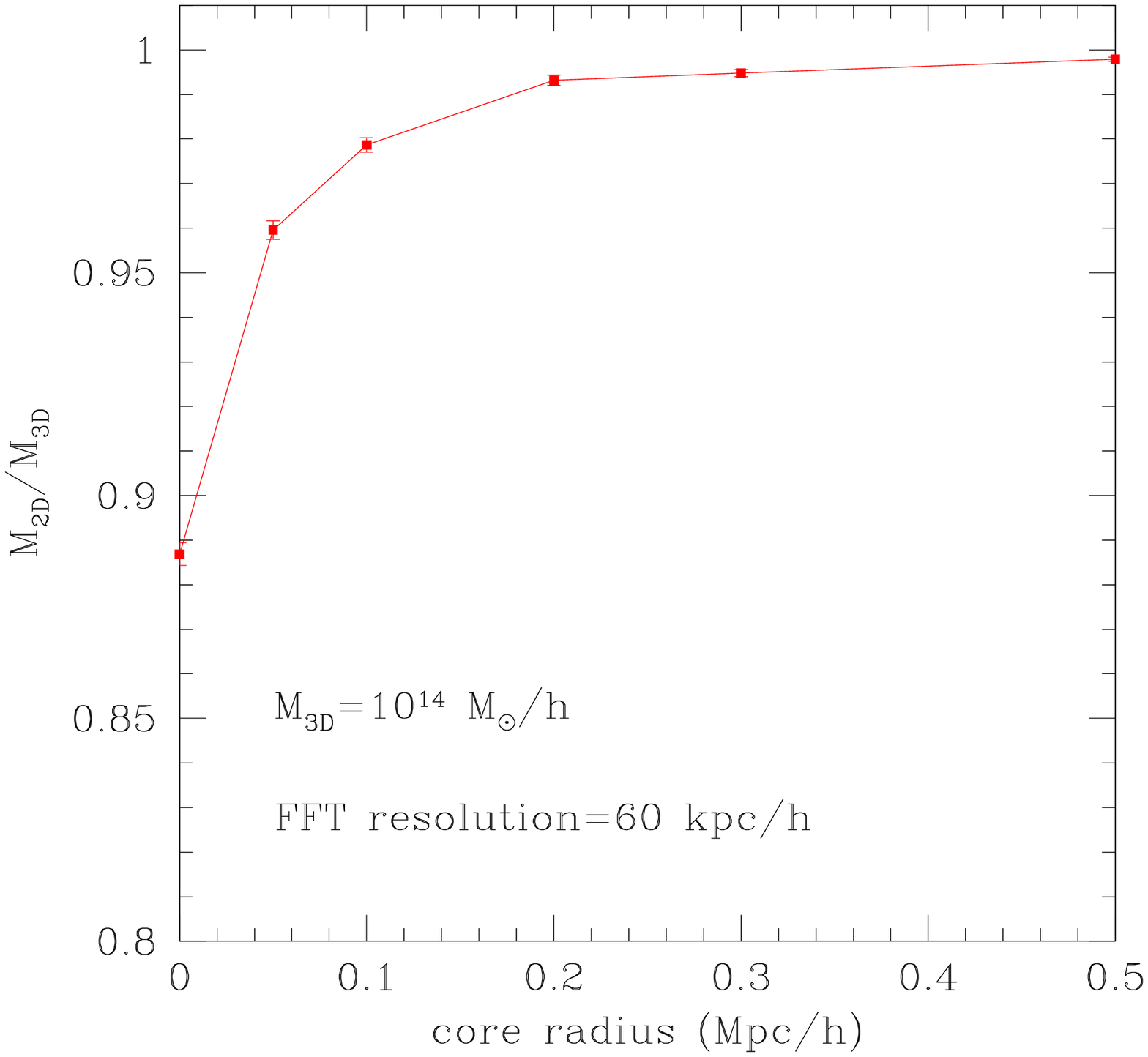}
\caption {The dependence of the filter efficiency on the core radius,
  for the Fourier mesh used throughout the paper (60 $\kpc$). The
  efficiency is defined as $M_{\rm 2D}/M_{\rm 3D}$, where $M_{\rm 3D}$
  is the Sheth-Tormen virial mass of the synthetic halo, and $M_{\rm
    2D}$ is the mass of the projected peak, assigned as described in
  \S\ref{Filter}. We show the average of measurements of 1000
  projected synthetic Sheth-Tormen halos, in the fiducial
  cosmology. We chose a core radius of 200 $\kpc$ to perform our
  analysis.}
\label{fig:CORE}
\end{minipage}
\hspace{0.6cm}
\begin{minipage}[t]{0.45\linewidth} 
\includegraphics[scale=0.4]{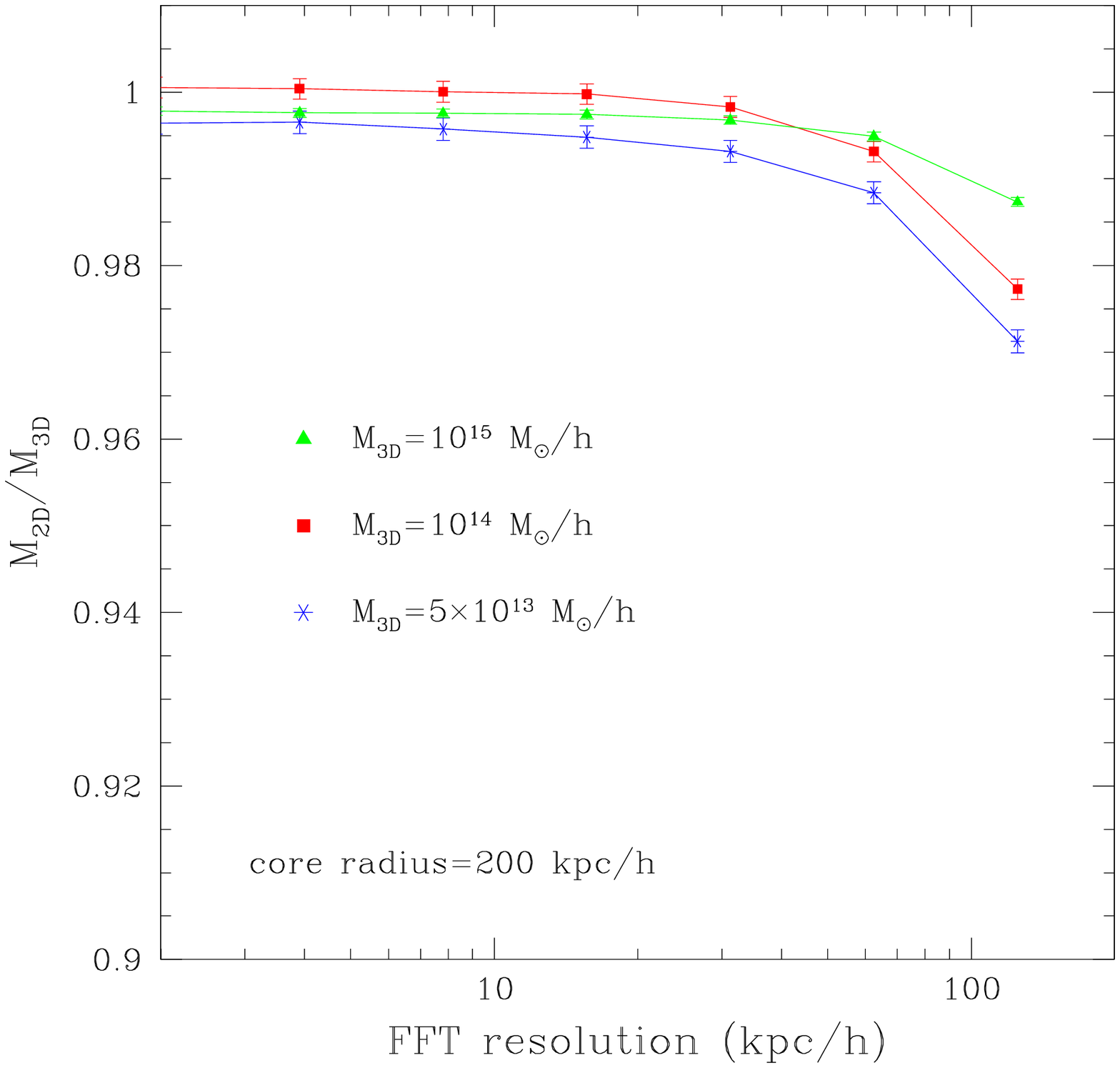}
\caption {The dependence of the filter efficiency on the Fourier mesh,
  for a core radius of 200 $\kpc$. The filter efficiency is defined as
  in Figure~\ref{fig:CORE}. For each mass we plot the average of the
  measurements of 1000 synthetic Sheth-Tormen halos in the fiducial
  cosmology. Throughout our analysis, we used a mesh of 60 $\kpc$.}
\label{fig:FFTRES}
\end{minipage}
\hspace{1.cm}
\end{figure*}
\par A nontrivial issue regarding the slabs is that of the peaks which
are situated right on the boundary of two slabs and are therefore
split in two. The consequence of such a boundary split is that the
peak is detected in neither slab or, even if it is detected in one of
them, its filter-associated mass is significantly reduced compared to
the true value. On average, in a 50 $\Mpc$ slab, about 20\% of the
peaks suffer from boundary-split effects. We addressed this problem by
considering two sets of slabs: `normal', and `interlaced'
slabs. Compared to the normal slabs, the interlaced ones are shifted
by half a slab thickness on the axis of projection. Thus, the boundary
of two normal adjacent slabs goes right through the middle of an
interlaced slab, so that halos that are divided in the normal slabs
appear whole in the interlaced slabs, and vice versa. We analyze the
peaks in both sets of slabs and compare them based on their location
in the $\hat{x}-\hat{y}$ plane (we consider the projection axis as the
$\hat{z}$ direction), and their filter-associated mass. To be precise,
the peaks in every normal slab are compared to those in the two
adjacent interlaced slabs which contain the normal slab. The peaks in
every interlaced slab are also compared to those in the two adjacent
normal slabs which contain the interlaced slab. If we find peaks with
the same or nearly the same $\hat{x}-\hat{y}$ coordinates, but with
masses that differ by more than 5\%, we consider that we have a
boundary split case and select the peak that has the higher mass. If
the difference in mass is smaller than 5\%, we keep the original
masses. 
\par Halos often `live' in filaments; if the filaments are oriented along
the projection axis, they can contribute to the mass of peaks sourced
by the respective halos, or they can even form peaks of their
own. Even when a halo is not right on a slab boundary, but it is close
to it, its filament can extend beyond the boundary. By splitting the
filament, the projected mass of the halo is slightly decreased. The
5\% threshold mentioned above not only removes the problem of split
halos, but it also makes split filaments less likely. For the
thickest slab with potential boundary problems, i.e. 256 $\Mpc$, we
use a higher threshold of 15\%, in order to avoid biasing our mass
measurements to higher masses. In this case there is only one
boundary, and hence fewer split objects.
\section{Testing the filter}
\begin{figure*}
\centering
\includegraphics[scale=0.6]{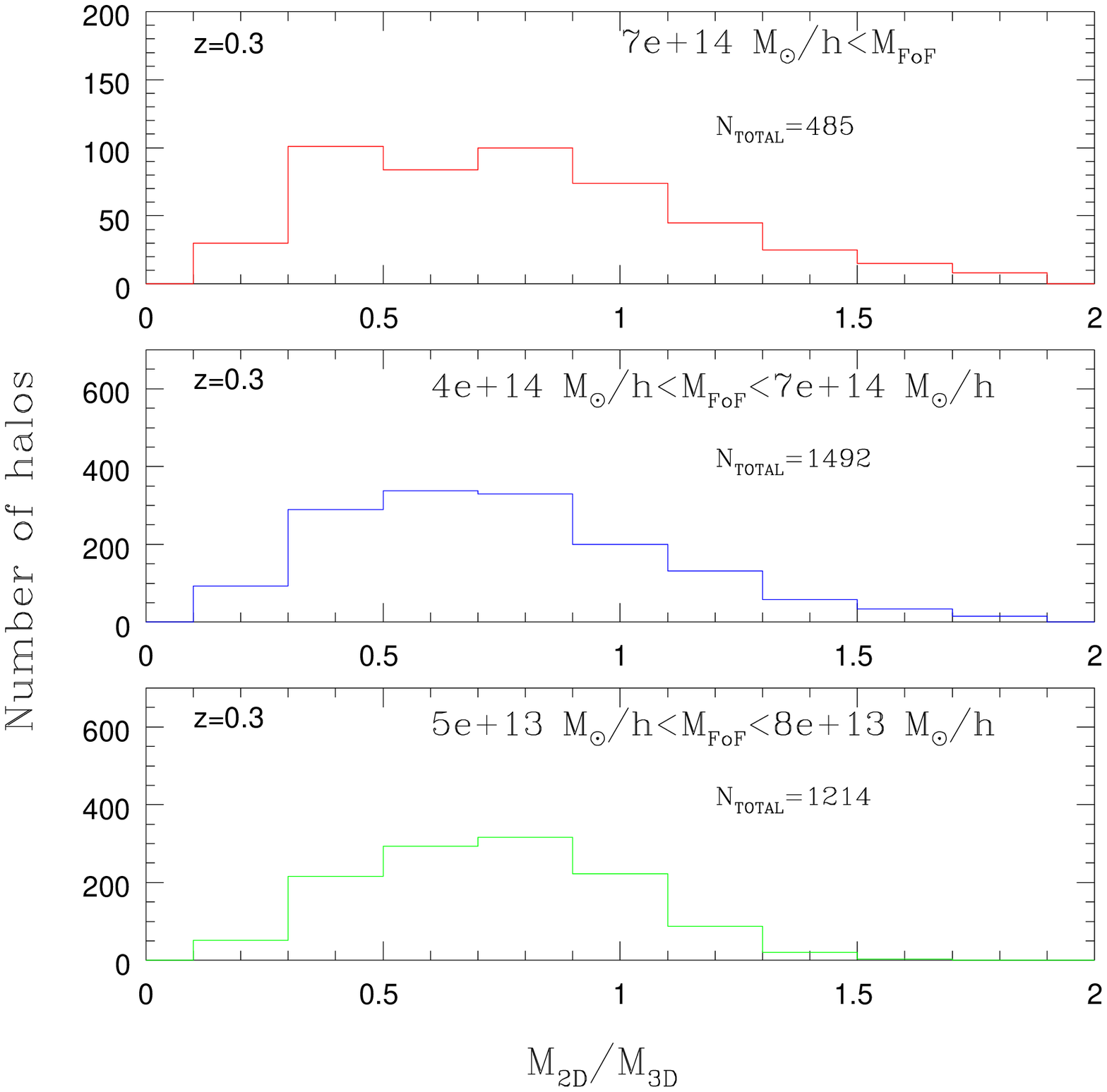}
\caption {The filter efficiency in the absence of projections. $M_{\rm
    3D}$ is the FoF mass of the simulation halos, and $M_{\rm 2D}$ is
  the mass of the peaks obtained by projecting the FoF halos by
  themselves, i.e. only the particles forming them. Three FoF mass
  bins are considered, with the total number of halos per bin
  indicated in the figure. We attribute the shift towards lower masses
  to departures of the FoF halos from the spherically-symmetric NFW
  profile. The dispersion in the recovered mass decreases for the bins
  of lower mass, in accord with the fact that the largest halos are
  the most triaxial ones.}
\label{fig:FOF_PEAK_HISTO}
\end{figure*}
\subsection{Numerical tests of the filter}
\label{Tests_synth}
The filter described earlier is optimal in the ideal conditions of
infinite particle-mass resolution. The fact that the simulation halos
have a discrete mass distribution changes the performance of any
filter when applied to simulation data. Also, our filter is tuned to
spherically symmetric halos, but real halos are more triaxial than
spherical.  We have investigated the efficiency of our filter in two
idealized cases: first, on synthetic NFW halos, to understand the
effects of the finite mass and mesh resolutions; second, on isolated
FoF halos (that is, only the particles from a single halo) of our
simulations, to see the effects of the halo asphericity. Both tests
were carried out for the fiducial cosmological model, as there is no
reason to assume that such numerical effects would manifest themselves
differently in the variational cosmologies.

\par The synthetic NFW halos that we have generated have the same mass
resolution as our simulations. The particles are sampled in the radial
direction with the NFW density profile, and their angular distribution
is uniform random on the sphere. The tests were performed for 3
different masses, using an ensemble of 1000 synthetic halos for each
mass. We found that the finite particle mass resolution mostly affects
the signal in the inner region of the clusters, defined by the scale
radius. That is because the NFW density profile has a significant
fraction of its mass enclosed within the scale radius ($\approx$
20\%).
\begin{figure*}
\centering
\includegraphics[scale=0.85]{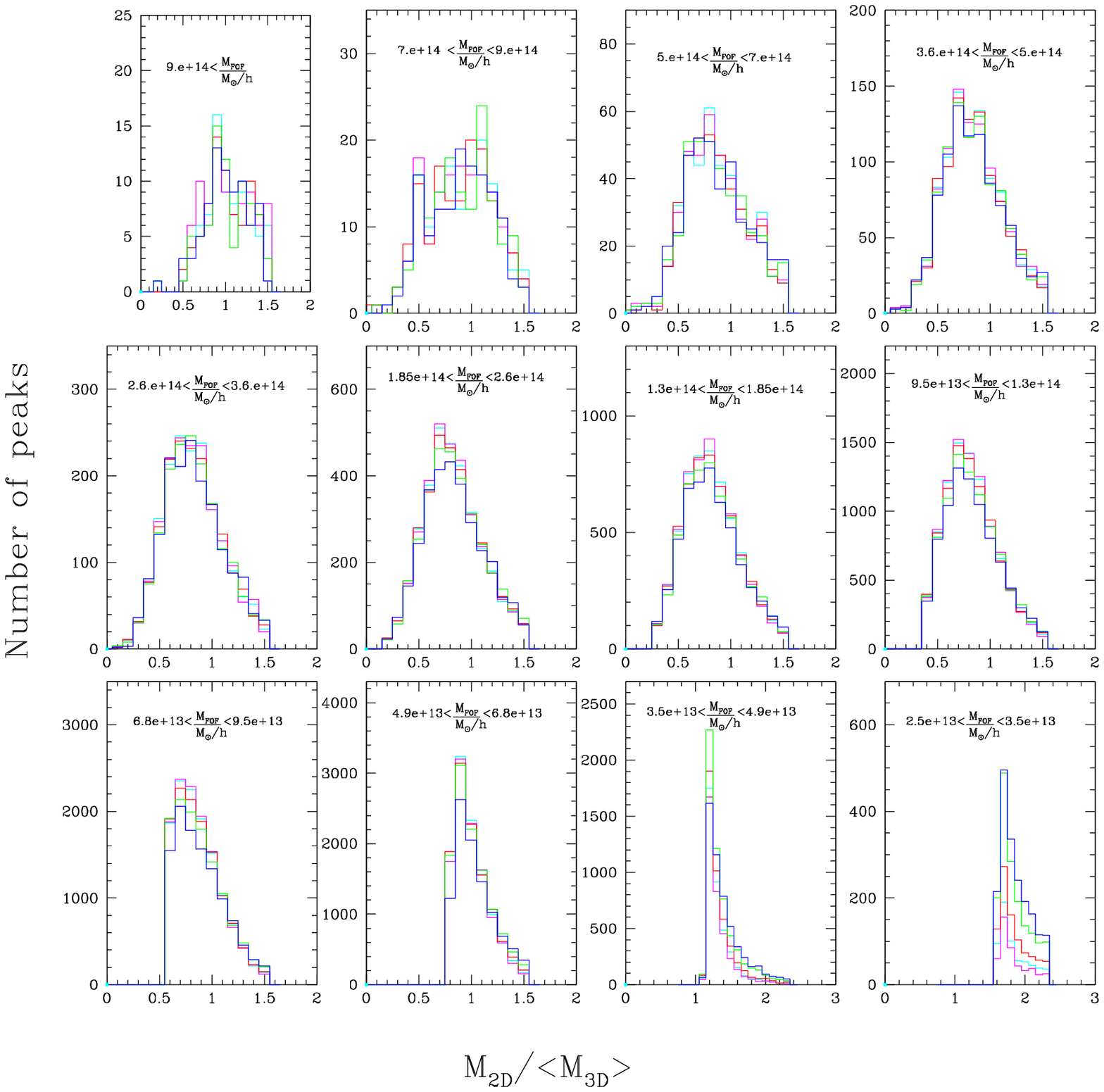}
\caption{The filter efficiency in the presence of projections. The
  efficiency is defined as $M_{\rm 2D}/M_{\rm 3D}$, where $M_{\rm 3D}$
  is the mass of the FoF halos that source the peaks measured in
  different slabs. $M_{\rm 2D}$ is the mass of the peaks formed by the
  FoF halos, when projected together with the rest of the matter in
  the slabs: 26 $\Mpc$ -- magenta-- , 51 $\Mpc$ --cyan-- , 102 $\Mpc$
  -- red -- , 256 $\Mpc$ -- green -- , 512 $\Mpc$ -- blue. The
  distribution is non-gaussian, and the spread is mostly due to
  departures of the 3D halos from the spherically-symmetric NFW
  profile.}
\label{fig:FOF_PEAK_SLABS}
\end{figure*}
Given the mass resolution of our simulations, the inner region of
average clusters is poorly populated with particles, and actually
unresolved for radii smaller than the softening length of the
simulations (60 $\kpc$, in our case). Therefore, we have chosen to
`core' the ST profile, i.e. we consider the density profile of the
filter to be constant for a radius $<r_{\rm core}$:
\be
\Sigma_{\rm \sss ST}(r)= \left\{ \begin{array}{lcl}
\Sigma_{\rm \sss ST}(r_{\rm core})\,, \hspace{0.5cm} r \leq r_{\rm core} \\
\Sigma_{\rm \sss ST}(r)\,, \hspace{0.5cm} r>r_{\rm core}
\end{array}\right.
\label{eq:core}
\ee
Thus, the traditional central cusp of the ST profile has been removed
from the filter, and the signal of the filtered halos is redistributed
so that the outer layers contribute more to it.  Even if our
simulations had allowed us to use the ST profile without removing its
cusp, there is a more physical reason that justifies our choice. Halos
relevant to WL surveys have an Einstein radius which is only 2-3 times
smaller than the scale radius. Since we are interested in the WL
regime, the measured signals of the halos must come from regions well
outside the Einstein radius, or else the WL approximation breaks
down. Note that the filtered density will be estimated from real data
by applying a corresponding filter to the shear map. The shear filter
corresponding to our constant-core ST profile has a `hole',
i.e. discards the signal coming from within the core radius. This is
necessary in practice since it is very difficult to measure the shear
around clusters at small radii from the center: most shear measurement
methodologies fail in regions of strong lensing, and cluster member
galaxies often obscure the lensed background galaxies.

Figure~\ref{fig:CORE} plots the dependence of the filter efficiency --
defined as $M_{\rm 2D}/M_{\rm 3D}$ -- on the filter core radius for an
ensemble of synthetic halos of mass $10^{14}\,\Msol$. $M_{\rm 2D}$ is
the mass that we assign to the density peaks in the way described in
section \S\ref{Filter}, and $M_{\rm 3D}$ is simply the ST virial mass
of the synthetic halos. The figure clearly shows the filter efficiency
to be $\approx$ 90\% even for halos obeying perfectly the NFW profile,
if no coring is done. Though not shown in the figure, we have also
probed that the efficiency would be more than 95\% if the mass of the
simulation particles were lower by a factor of 10, with no core
applied. However, since this mass resolution is not available,
throughout this study we have adopted a core value of 200 $\kpc$. This
is also valid for the results published in
\citet{2009ApJ...698L..33M}.
\par Equation~(\ref{eq:fil1}) is a real space convolution. By
smoothing the density field of dark matter with a filter, we expect to
erase the structure on scales smaller than the scale of the filter. It
is convenient to do the smoothing in Fourier space, and in order to
compute the discrete Fourier transform of the density field, we
project it on a grid. We refer to the ratio of the simulation cube
side $L$ and the number of grid points taken on that side, as the
Fourier mesh resolution. Our whole analysis of the simulation data was
done with a mesh resolution of 60 $\kpc$. In Figure~\ref{fig:FFTRES} we
plot the filter efficiency dependence on the Fourier mesh resolution
for the chosen core value. Large halos are less affected than small
halos by resolution effects, and in general the filter efficiency is
of about 98\% for our mesh and core chosen values. Since we do not
wish to include in our analysis halos that are too small and too
`poor' in particles, we have imposed a mass detection threshold of
$5\times 10^{13}\,\Msol$ for the projected peaks. We do not
include in our analysis peaks with mass below this threshold.

\par Figure~\ref{fig:FOF_PEAK_HISTO} shows the correspondence of FoF
masses and 2D masses recovered by our filter, {\em in the absence} of
projections. The FoF halos have been projected by themselves: only the
particles identified by the halo finder as being part of a halo have
been projected. Again, the efficiency is defined as $M_{\rm 2D}/M_{\rm
  3D}$: $M_{\rm 2D}$ is the mass assigned to the density peaks
according to section \S\ref{Filter}, and $M_{\rm 3D}$ is the FoF mass
of the simulation halos that we analyze. The filter efficiency is
depicted for 3 large mass bins. The FoF halos were selected from all
eight realizations of the fiducial cosmology. The plot reveals that
the vast majority of the FoF halos are shifted to lower masses by our
filter. This is mostly due to departures of the halo density profiles
from the spherically symmetric NFW profile: the filter given by
Eq.~(\ref{eq:fil2}) is a spherical overdensity (SO) filter that we
apply to FoF halos. FoF and SO masses are known to differ. Indeed, had
we applied an SO 3D halo finder, or had we used another FoF linking
length, the filtered masses would have had a different distribution
and mapping to the 3D halos. It is also possible that the scatter in
the concentration parameter, which is well known to be described by a
log-normal distribution (\citet{2001MNRAS.321..559B}) might also cause
some of the scatter seen in Figure~\ref{fig:FOF_PEAK_HISTO}. Also,
halos have substructures, e.g. \cite{1999ApJ...524L..19M}. Since we
are interested in the cosmology scaling of the projected-peak mass
function, and not in the analysis of individual clusters, the mapping
between FoF halo masses and filtered masses is not, in this study, of
great significance to us. After all, there is no absolute way to
define a halo mass.
\begin{figure*}
\centering
\includegraphics[scale=0.65]{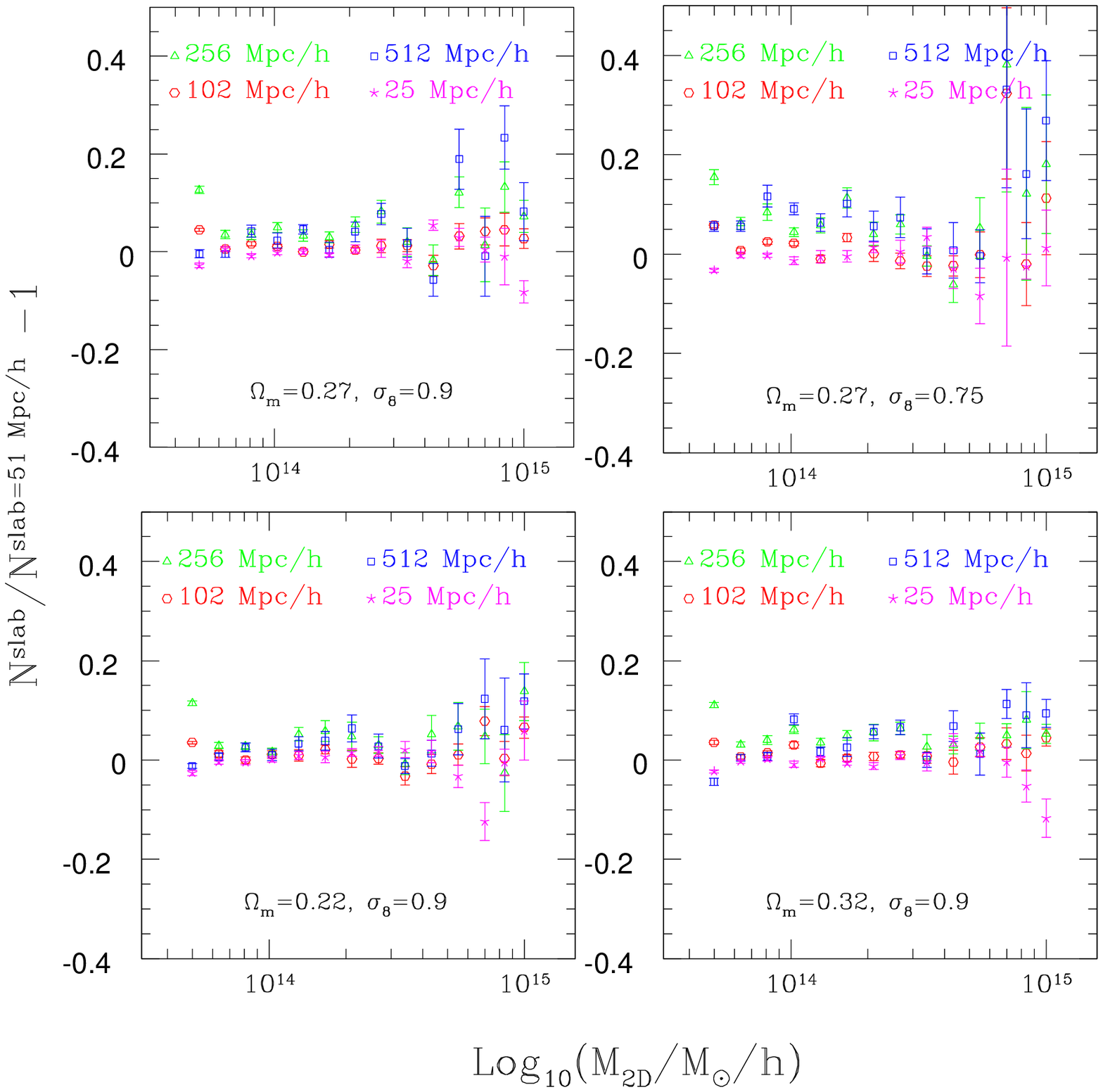}
\caption {The fractional difference of the number of projected peaks
  in slabs of 25, 102, 256, 512 $\Mpc$ and the peaks in the fiducial
  slab of 51 $\Mpc$. Each panel corresponds to one cosmology. The
  difference in the mass functions for different slabs is due to the
  correlated projected term described by Eq.~(\ref{eq:M_corr}), as
  well as to uncorrelated projections in the case of the thickest
  slabs.}
\label{fig:UC_FRAC_MF}
\end{figure*}
It is worth emphasizing that, unlike \citet{2001ApJ...547..560M} and
\citet{2004MNRAS.350.1038C} our recovered 2D masses are not the result
of averaging projections of the same clusters along different lines of
sight. The halos are projected on the same axis as the simulation
cubes, and to each of them we apply our filter `blindly', i.e. without
assuming any prior knowledge of the true position of the cluster
center. This is valid for our whole analysis. Had our individual
cluster mass estimates been an average over several lines of sight, we
expect to have found a reduced scatter and a smaller skewness in the
distributions shown in Figure~\ref{fig:FOF_PEAK_HISTO}.

\par Figure~\ref{fig:FOF_PEAK_SLABS} shows the correspondence of FoF
masses and 2D masses in {\em the presence} of projections. Here we
have matched the FoF halos of the simulations with the peaks projected
in slabs of various thickness. The different colors indicate the
thickness of the projection slabs. The binning of the FoF masses is
finer than in the previous figure. The sharp cut-off visible in the
right lowest panels stems from our selection of peaks larger than
$5\times 10^{13}\,\Msol$. The halos sourcing these peaks can however
have masses lower than this threshold. Not surprisingly, the same
trend present in Figure~\ref{fig:FOF_PEAK_HISTO} is also visible
here. Again the FoF halos are shifted towards lower masses, and the
distribution of peaks is non-gaussian. A log-normal distribution fits
some of the intermediate mass bins. The slab thickness does not seem
to play a significant role in shaping the distribution. While in this
study Figure~\ref{fig:FOF_PEAK_SLABS} is just illustrative, should we
wish to predict the projected-peak mass function from the 3D one, we
would need the distribution $M_{\rm 2D}-M_{\rm 3D}$, as suggested by
\citet{2001ApJ...547..560M}.
\begin{figure*}
\centering \includegraphics[scale=0.65]{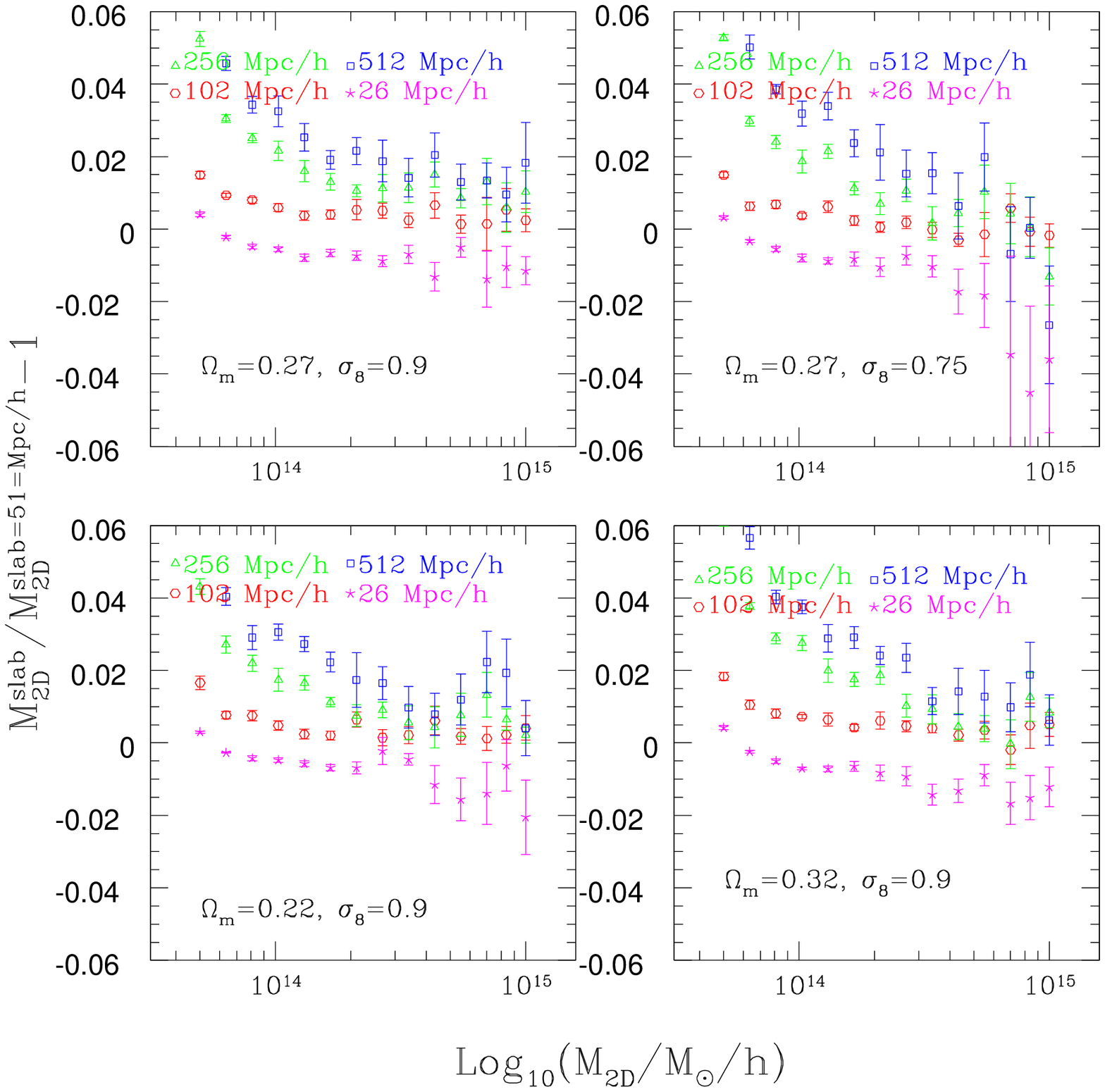}
\caption {The fractional change in mass due to correlated projections
  for peaks measured in different slabs and the fiducial slab
  peaks. Every peak projected in slabs with the fiducial thickness,
  was then tracked among the peaks projected in the other slabs. }
\label{fig:CORR_MASS_CHANGE}
\end{figure*}
Figures~\ref{fig:FOF_PEAK_HISTO} and~\ref{fig:FOF_PEAK_SLABS} serve as a
warning against blind comparisons between 3D and 2D quantities, as
often seen in the literature. Figure~\ref{fig:FOF_PEAK_HISTO} is an
example of `filter effects', i.e. scatter between 2D and 3D masses due
to different mass definitions and other causes, but completely
unrelated to LSS projections. Figure~\ref{fig:FOF_PEAK_SLABS} shows
that such filter effects are dominant compared to the correlated
projection effect. Unless the 2D and 3D peaks are found with {\em the
  same} filter, it is very difficult to measure the impact of
correlated projections directly on 3D masses. In order to remove the
filter effects shown in Figure~\ref{fig:FOF_PEAK_HISTO}, one should
measure the change in the mass of peaks projected in slabs of
different thickness. This is how we proceed in the following sections.
\section{RESULTS}
\label{Results}
In our previous study \citet{2009ApJ...698L..33M}, we analyzed the
effect of correlated projections in slabs of 51 $\Mpc$. Here we extend
our analysis to several slab thicknesses: 26 $\Mpc$, 102 $\Mpc$, 256
$\Mpc$, and 512 $\Mpc$.
\subsection{Slab thickness tests}
\begin{figure}
\centering \includegraphics[scale=0.4]{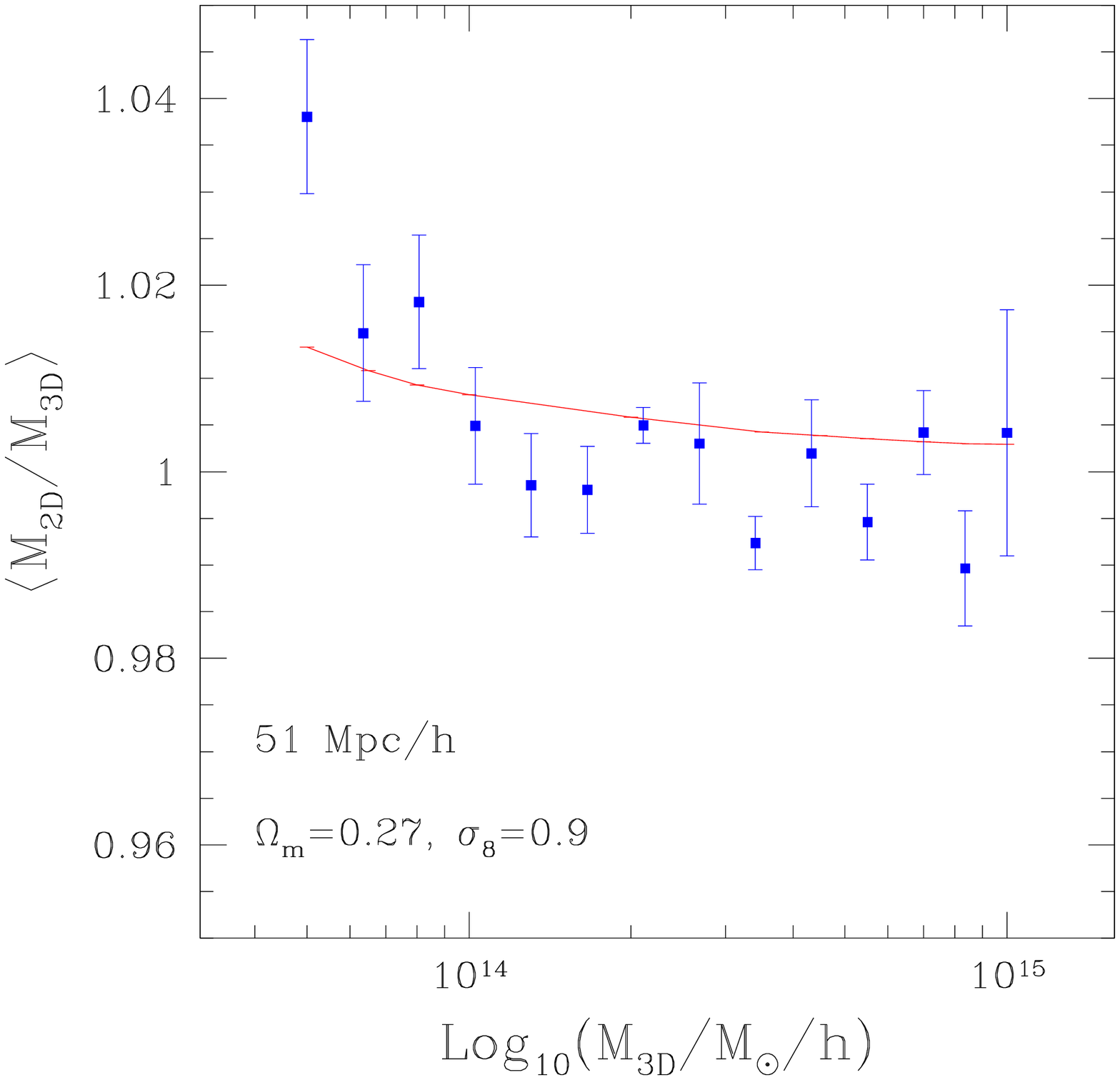}
\caption {A test of Eq.~(\ref{eq:M_corr}): $M_{\rm 3D}$ is the mass of
  FoF halos projected by themselves, and filtered according to
  Eq.~(\ref{eq:fil2}). $M_{\rm 2D}$ is the mass of the peaks sourced by
  the same FoF halos when projected in slabs of the fiducial
  thickness, together with all the mass in the respective slabs. The
  solid line is the halo model prediction from Eq.~(\ref{eq:M_corr}).}
\label{fig:2D_3D_CHANGE}
\end{figure}
\begin{figure*}
\centering \includegraphics[scale=0.65]{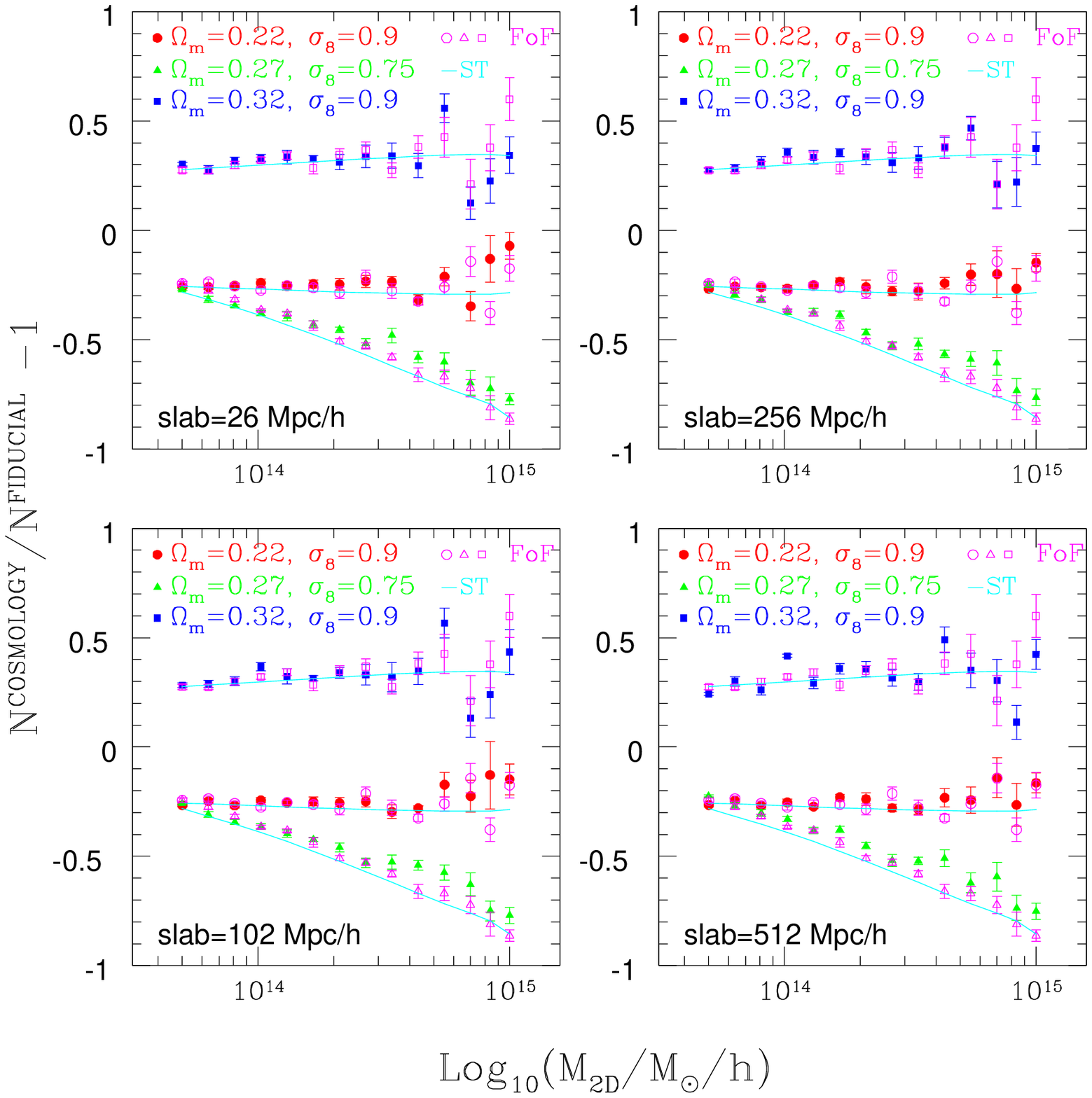}
\caption {The mass function of the variational cosmologies, scaled by
  fiducial cosmology mass function, considering slabs of different
  thickness. Filled symbols are peak measurements, open ones are FoF
  halos, and the continuous lines indicate the Sheth-Tormen mass
  function. The error bars represent errors on the mean of 8
  realizations of each cosmology.}
\label{fig:FRAC_MF}
\end{figure*}
\par First we probe the dependence of the projected density peak
abundance on the slab thickness. 
Figure~\ref{fig:UC_FRAC_MF} shows the fractional difference of the
peak number counts in the slabs that we investigate, compared to our
`fiducial' slab of 51 $\Mpc$. We notice the fractional differences
deviate from 0, particularly at the high-mass end, an anticipated
effect of projections. The deviations depend on the slab thickness and
also on the cosmological model -- the low-$\sigma_{8}$ cosmology shows
greater sensitivity to the slab variations than the other three
models. We expect the most significant changes from the fiducial slab
peak abundance to occur precisely at the high-mass end: there are few
high-mass clusters, so a small variation in their mass can alter their
distribution on the exponential tail of the mass function
considerably. In particular, for the 512 $\Mpc$ and even the 256
$\Mpc$ slabs, this variation may be increased by the presence of
chance projections along the line of sight, which seems to be
suggested by Figure~\ref{fig:UC_FRAC_MF}. The correlation length of
large clusters is $\approx 30\,\Mpc$, therefore we do not expect
correlated projections to affect the abundance of peaks in these
slabs. There is also an increase in the high-mass peak abundance
corresponding to the fiducial slab compared to the 26 $\Mpc$ one,
while only a very small rise is visible in the 102 $\Mpc$ slab
compared to the fiducial one---a sign that the correlation function of
clusters drops significantly at scales larger than 50 $\Mpc$.
\begin{figure*}
\centering
\begin{minipage}[t]{0.45\linewidth} 
\includegraphics[scale=0.4]{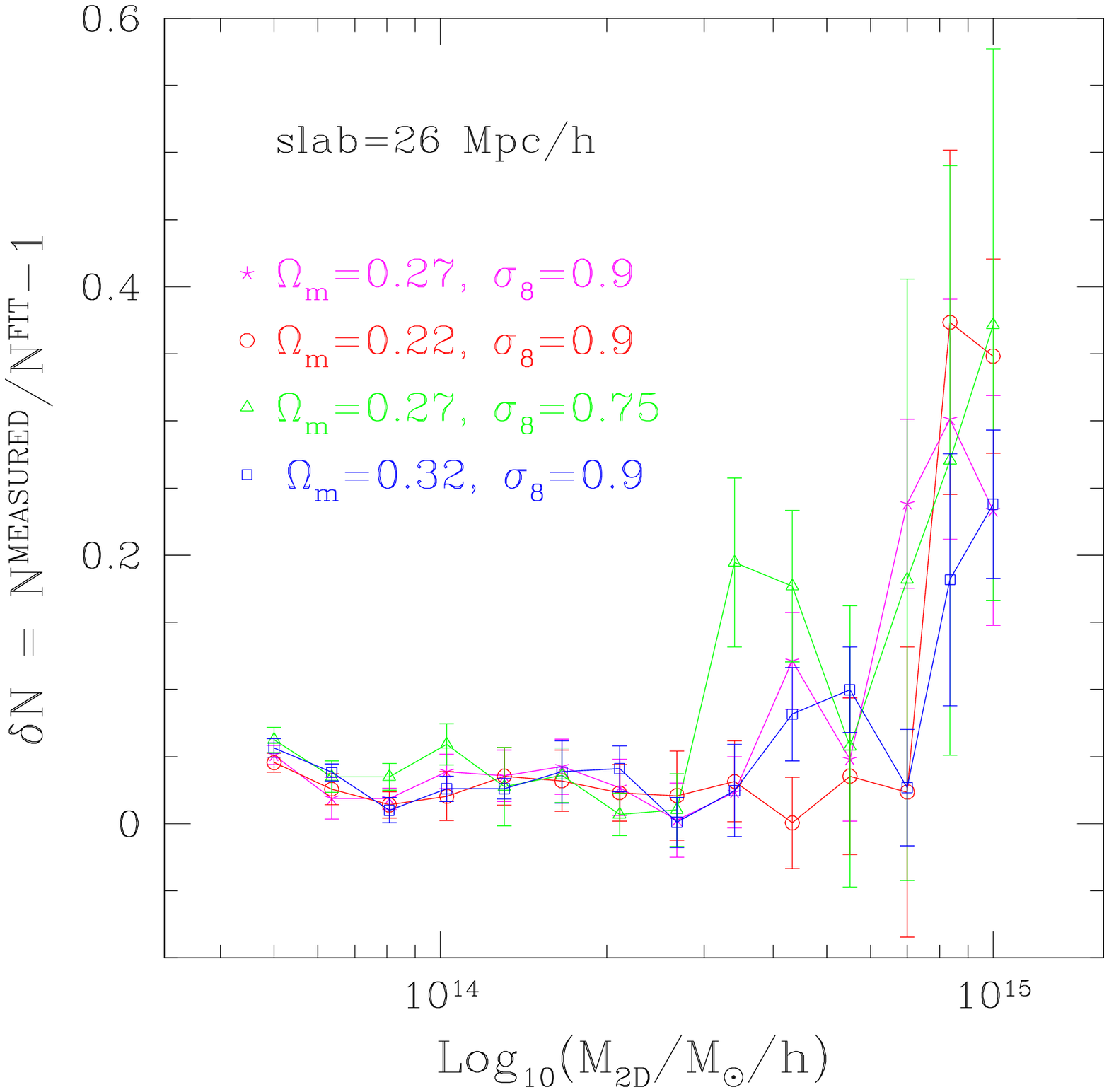}
\caption {Fractional difference of measured and fitted 2D mass
  functions for all four cosmologies. The slab thickness is 26 $\Mpc$.}
\label{fig:DATA_FITS_NS20}
\end{minipage}
\hspace{1.cm}
\begin{minipage}[t]{0.45\linewidth}
\includegraphics[scale=0.4]{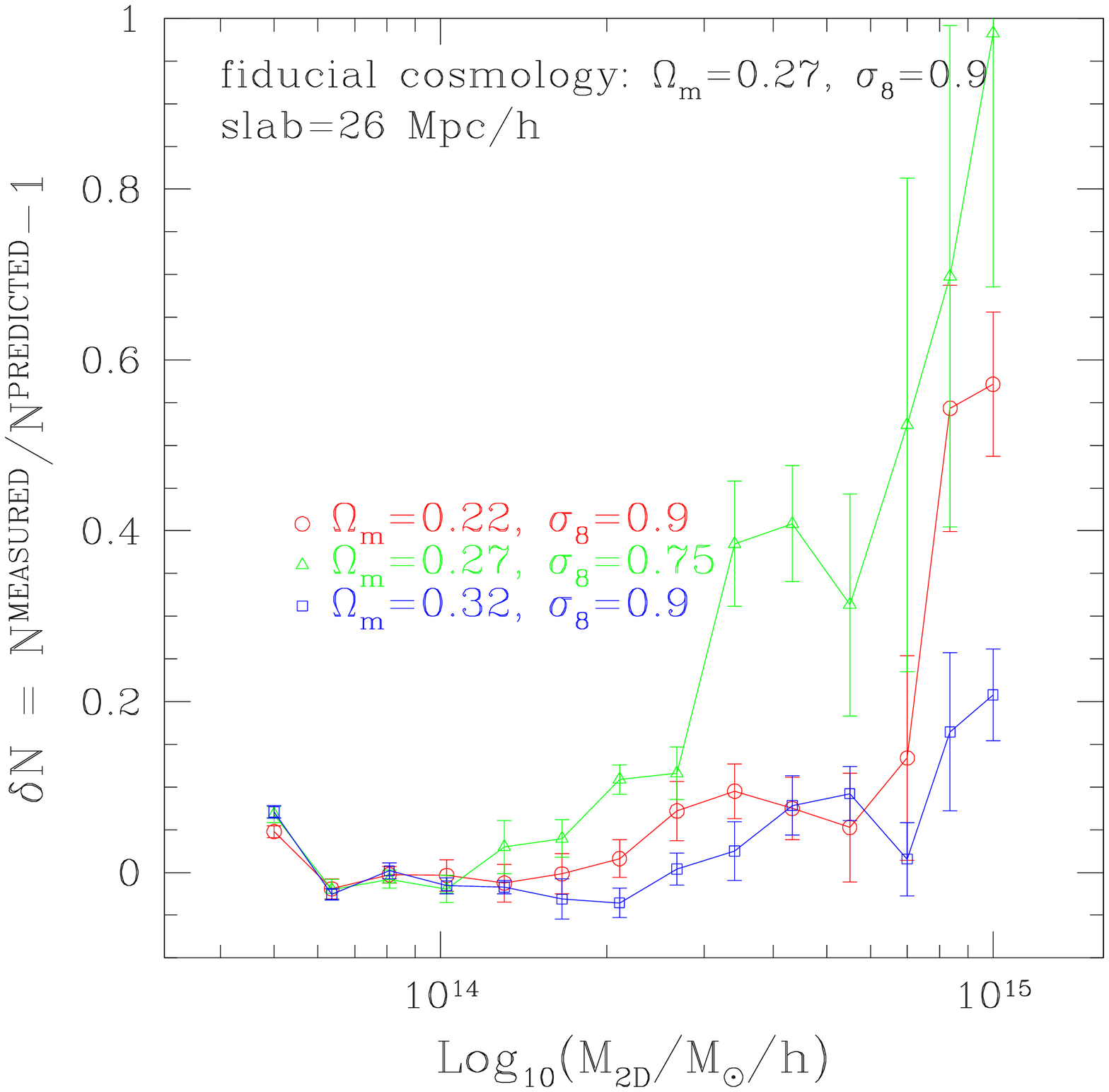}
\caption{Fractional difference of the 2D mass function measured in
  slabs of 26 $\Mpc$ and the prediction given by Eq~(\ref{eq:fit2}).}
\label{fig:RATIO_FIT_ST_NS20}
\end{minipage}
\end{figure*}
Overall, we are satisfied with the robustness of the mass function
projected in slabs of various thickness: analyzing the projected
density field with a compensated filter has rendered the resulting
peak abundance largely independent of the slab thickness.
\subsection{The mass change induced by correlated projections 
on individual peaks}
In section~\S\ref{HMT} we used the halo model to predict the change in
the projected mass of a cluster induced by structures correlated with
it, i.e. Eq.~(\ref{eq:M_corr}). We have evaluated this expression for
the particular case of our filter, described by
Eq.~(\ref{eq:fil2}). In this case, Figure~\ref{fig:HM_PRED} indicated
that there should be a very small change ($< 2\%$) in the 2D mass of
peaks -- defined in section~\S\ref{Filter} -- , compared to their 3D
mass -- defined as the ST virial mass. In this section we test the
halo model prediction. Since our 3D masses have not been identified
with an SO filter, but with the FoF algorithm, we cannot probe the
result of Eq.~(\ref{eq:M_corr}) directly. Instead, we circumvent this
difficulty by comparing only masses of objects identified with the
same filter.

First we follow the mass change for the simulation clusters when
projected in slabs of various thickness. We have traced every peak in
the fiducial slab among the peaks corresponding to the other slabs. We
have selected only those peaks that maintained roughly the same $\hat
x-\hat y$ coordinates in all the slabs where they were projected -- we
did not allow the coordinates to shift by more than
$\approx120\,\kpc$.  This restriction was imposed to ensure that we
compare the mass of \mx{\it{the same}} peaks traced in different
slabs. Figure~\ref{fig:CORR_MASS_CHANGE} presents the fractional
change in mass for the peaks projected in the variational slabs
compared to the fiducial slab (symbols). The error bars represent
errors on the mean of the eight realizations of each cosmology. We
have measurements for all four cosmologies. One remarkable feature may
be noted: the simulation data indicate the impact of correlated
projections on the measured halo masses to be \mx{\it{very small}},
i.e. about 2\% for peaks measured in slabs of 26 $\Mpc$ and 102 $\Mpc$
relative to the fiducial slab peaks. The exception is the
low-$\sigma_8$ cosmology, which shows stronger deviations at the
high-mass end. The halo model prediction for the change in mass of
peaks projected in slabs of 26 $\Mpc$ and 102 $\Mpc$ compared to peaks
projected in slabs of 51 $\Mpc$ is that it should be smaller than 0.1
\%. Figure~\ref{fig:CORR_MASS_CHANGE} indicates a more noticeable
change, however. The slightly larger measured mass changes (~4\%) for
the two thickest slabs are likely due to chance projections.

Second, we carry out a comparison between peaks measured in slabs of
thickness 0 and peaks measured in the fiducial slab. To be precise, we
take a large number of the FoF halos ($\approx 8000$) from the
fiducial cosmology, and project only the particles within the halos
themselves. An identical procedure was followed to obtain the
distribution in Figure~ \ref{fig:FOF_PEAK_HISTO}. In this case, we
filter a larger number of halos, and take finer mass bins. We consider
the mass of the peaks obtained through the projection of the FoF halos
by themselves as a proxy for the 3D ST virial mass. We then compare
this mass to the mass of the peaks sourced by the same FoF halos when
projected together with all the matter in slabs of the fiducial
thickness. We do not expect this approximation to be perfect, but it
is the closest measurement we can make on our data for a direct
comparison with the halo model prediction shown in
Figure~\ref{fig:HM_PRED}. The result is shown in
Figure~\ref{fig:2D_3D_CHANGE}. Some caution is necessary when
interpreting this figure, as for some of the intermediate mass bins,
the ratio $M_{\rm 2D}/M_{\rm 3D}$ goes below unity. Naturally, this
should not be the case. We explain this anomaly by the fact that when
an FoF halo is projected together with the structure in which it is
embedded, the surrounding matter renders it more homogeneous and the
compensated filter may therefore attribute it a smaller mass than when
the halo is projected by itself. However, despite this anomaly,
Figure~\ref{fig:2D_3D_CHANGE} should be taken as supportive of the
halo model prediction, i.e. Eq.~(\ref{eq:M_corr}). Hence we expect
that correlated projections alter the masses of clusters at the
percent level, and therefore are not a dominant systematic for WL
cluster mass measurements. If there is a projection bias in WL mass
determinations, it is more likely due to structures uncorrelated with
the main lens.
\subsection{Predicting the projected mass function}
We now try to address two of the main objectives of this work: in the
context of correlated projections, we examine the behaviour of the 2D
mass function and also the possibility to predict this behaviour. In
\citet{2009ApJ...698L..33M} we have shown that the scaling with
cosmology of the 2D mass function projected in slabs of 51 $\Mpc$
follows very closely the scaling of the 3D mass function and also that
of the semi-analytical ST model.

Figure~\ref{fig:FRAC_MF} is similar to Figure 2 from
\citet{2009ApJ...698L..33M}, but generalizes that result to other slab
thicknesses. The figure shows the fractional difference of the peak
counts for the variational cosmologies scaled by the fiducial
cosmology peak counts (solid points). It also shows the same
fractional differences for the 3D FoF mass functions measured from the
simulations (open magenta symbols), as well as the ST predictions
(solid lines). The error bars are on the mean of the eight
realizations of each cosmology, and the cosmic variance between
realizations is minimized by the choice of initial conditions
mentioned in section \S\ref{Sims}. The difference in the SO filter and
FoF mass definitions is also minimized: we compare 2D peaks with 2D
peaks and FoF halos with FoF halos.

We maintain the conclusion of \citet{2009ApJ...698L..33M} regarding the
very similar scaling with cosmology of the 2D, 3D, and ST mass
functions.  


We next try to fit the lensing mass function corresponding to the slab
thickness of 26 $\Mpc$. We shall assume that the functional form
proposed by \citep{1999MNRAS.308..119S} can be used to describe the
projected mass function, but with the parameters of the model
recalibrated. The ST mass function is given by:
\be \frac{dn_{\rm \sss ST}(M)}{d\log M}\,d\log M = \frac{\rho_{\rm m}}{M} f_{\rm \sss ST}(\nu)
d\nu \,,
\ee
with $\nu=\delta_{\rm sc}^{2}/\sigma^{2}$ and:
\be
f_{\rm \sss ST}(\nu)=A\sqrt\frac{2 a \nu}{\pi}\left[1+(a\,\nu)^{-p} \right]\exp\left(-a \nu/2\right).
\label{STmf}
\ee
In the above $dn_{\rm \sss ST}=n_{\rm \sss ST}(M)\,dM$ gives the
number density of halos, $\sigma^2$ is the variance of the
top-hat-smoothed linear density field, $\delta_{\rm sc}=1.69$ is the
critical density for spherical collapse. The ST model has two free
parameters, with the following values: $a=0.707$, $p=0.3$. The
normalization parameter $A$ is obtained from the constraint that all
mass in the Universe is in halos:
\be\int_{0}^{\infty} dM M n_{\rm \sss ST}(M) = \bar{\rho} .\ee
This leads to:
$A(p)=\left[1+2^{-p}/\sqrt\pi\;\Gamma(1/2-p)\right]^{-1}$. For the ST
mass function, $A=0.3222$. The projected mass function fit conserves
this normalization relation, but changes the values of $a,\,p$. To fit
the parameters we follow the approach of \citet{2006ApJ...646..881W}
and compute the extended Poisson likelihood for our simulations:
\be 
\ln\lambda(\mx{\bom $q$})=-\sum_{j=1}^{N_{\rm r}}\sum_{i=1}^{N_{\rm
    m}}\left[\mu_{ij}(\mx{\bom $q$})-n_{ij}+n_{ij}\ln
  \frac{n_{ij}}{\mu_{ij}(\mx{\bom $q$})}\right], 
\ee
where $N_{\rm r}$ is the number of realizations per cosmology -- 8 in our
case -- , and $N_{\rm m}$ is the number of mass bins considered.  $n_{ij}$
is the measured number of projected peaks in mass bin $i$ of
realization $j$, while $\mu_{ij}$ is the predicted number of halos for
mass bin $i$ and realization $j$. We find the point in the parameter
space defined by $\mx{\bom $q$}$=$\left(a, p\right)$ which maximizes
the above likelihood, and thus we determine the best fit for the
projected mass function. Note that, had we been attempting to
constrain the mass function at lower masses, we expect that the
Poisson model would have not captured the true variance in the
distribution, owing to sample variance fluctuations
(\citet{2003ApJ...584..702H, 2009arXiv0907.0019C}).

We have fitted the free parameters for each cosmological model, and
our results are the following:
\begin{enumerate}
\item Model 1:  $a=0.719,\, p=0.298,\: A=0.3242$
\item Model 2:  $a=0.690,\, p=0.306,\: A=0.3161$
\item Model 3:  $a=0.665,\, p=0.318,\: A=0.3034$ 
\item Model 4:  $a=0.727,\, p=0.292,\: A=0.3301$
\end{enumerate}
Figure~\ref{fig:DATA_FITS_NS20} shows a comparison between the
measured and the fitted projected mass functions for all cosmologies.
We note that the ST functional form is not ideal to fit the high-mass
end of the 2D peak function: the accuracy of the fits is better than
10\% for the low and intermediate mass bins, but drops to 30\%-40\%
for the high mass bins. This is just a consequence of having
projection noise added to measurements on a steep mass function. We
also try to predict the projected mass functions of the variational
models by using only the fit to the fiducial cosmology and ratios of
the \mx{\it{standard}} ST mass functions:
\be
n^{X}=n_{\rm{fit}}^{\rm{fiducial}}\times\frac{n_{\rm \sss ST}^{X}}{n_{\rm \sss ST}^{\rm fiducial}}.
\label{eq:fit2}
\ee
In the above, $n^{X}$ is the predicted 2D mass function for any of the
variational cosmologies, $n_{\rm{fit}}^{\rm{fiducial}}$ is the fit for
the 2D mass function of the fiducial cosmology; $n_{\rm \sss ST}^{X}$
and $n_{\rm \sss ST}^{\rm fiducial}$ are the standard ST mass
functions for the variational and fiducial cosmologies respectively.

 Figure~\ref{fig:RATIO_FIT_ST_NS20}
presents the comparison between this prediction and the measured 2D
mass functions. The low- and high-$\Omega_m$ cosmology mass functions
are captured with an accuracy of 10\%-20\%, at low and intermediate
mass bins, while the low-$\sigma_8$ model is worse. The modeling fails
completely at the high-mass end, which was to be expected given that
for this mass range: (i) there is a large inaccuracy in the fit of the
fiducial cosmology; (ii) the ST functional form does not describe well
the 2D mass functions of the variational cosmologies either. This
simplistic modeling should just be taken as a proof of concept, that
the scaling with cosmology of the projected mass function presented in
Figure~\ref{fig:FRAC_MF} can be used as a starting point for more
complicated models of the 2D peak abundance.
\section{Conclusions}
\label{Conclusions}
Future or near-term WL surveys will be able to detect clusters as
peaks in shear maps. Projections from LSS are likely to make it
difficult to convert the amplitude of the shear peaks to
traditionally-defined cluster masses. However, it is possible to
circumvent this conversion if the cosmology dependence of the
shear-peak counts is understood. In this case, one can constrain
cosmology by measuring directly the shear-peak abundance. While this
can be done in dark matter simulations, as recently shown by
\citet{2009arXiv0906.3512D}, it is not practical to rely only on
numerical methods. The standard cosmological model contains 6 free
parameters, and so the space of models to be tested may quickly become
large. Thus, if a well-understood analytical framework can be
identified for the peak counts, it will hopefully facilitate an
accurate interpolation between expensive numerical predictions made in
this high-dimensional space of cosmological models. This is the main
motivation for our study.

In this paper we focused attention on estimating and understanding the
effect of `correlated' projections on the mass function of density
peaks projected in slabs of different thicknesses. We reserve
`uncorrelated' projections for future study.
Correlated projections are ideally studied in slabs, given the fact
that the correlation length of average clusters is $\approx 30\,
\Mpc$. In such slabs, the projected density is equal to the
convergence up to a constant, so it is perfectly adequate to choose it
as an observable. 

In \S\ref{Sims} we described the large ensemble of numerical
simulations that were generated for our study. These consisted of 4
cosmological models with 8 simulations per model and with total volume
per model $V_{\rm tot}=1 [{\Gpc}]^3$.  

In \S\ref{HMT} we presented a calculation based on the halo model of
structure formation of the expected change in the mass of 3D clusters
induced by correlated projections. We showed that the mass increase
depends on the filter applied to the projected density field. For a
top-hat filter, the increase in mass due to correlated projections is
of $10-15\%$. For a matched, optimal, compensated filter, it is
roughly a factor of 8 lower. Thus, the halo model predicts that
correlated projections should not alter the virial mass of clusters by
more than 2\%.

In \S\ref{Algorithm} we described our algorithm for identifying
projected-density peaks. One important new element was that we offered
a solution to the problem of `peaks-in-peaks'. We did this through
applying a hierarchical filtering strategy. We used different scales
for the matched filter and took a top-down approach to peak
detection. Smaller peaks that were contained within higher mass peaks
were discarded.

\par We have used the same optimal filter as in
\citet{2006PhRvD..73l3525M}. This filter is compensated, maximizes the
S/N, and matches the NFW density profile. Its normalization is chosen
so that it returns the ST virial mass of halos. Tests on synthetic
data, i.e. perfect NFW halos, have shown in section
\S\ref{Tests_synth} that the filter requires `coring' when applied to
simulated data, due to the finite mass and mesh resolutions. The
coring will also be necessary for application to real WL measurements,
due to practical difficulties of making precise WL measurements in the
inner regions of galaxy clusters.

We have measured the distribution $M_{\rm 2D}/M_{\rm 3D}$ -- where
$M_{\rm 3D}=M_{\rm FoF}$ -- in the presence of projections: we have
matched the FoF halos of the simulations and the peaks measured from
the density field projected in slabs of different thickness. We have
found this distribution scattered and biased towards values smaller
than 1. We have also measured the same distribution in the absence of
projections, by projecting the FoF halos by themselves. Even in this
case the distribution $M_{\rm 2D}/M_{\rm 3D}$ was scattered and biased
towards values smaller than 1. This shows that most of the scatter
between $M_{\rm 2D}$ and $M_{\rm 3D}$ is not due to LSS projections,
but rather to the departures of the FoF halos from the
spherically-symmetric NFW profile, to the existence of substructures,
and possibly to the scatter and stochasticity of the concentration
parameter.

The point of these tests is to clarify a somewhat loose perception in
the literature concerning the impact of LSS projections on cluster
mass measurements: when comparing 2D versus 3D quantities (i.e. mass,
S/N, etc.), one must account for the halo/cluster identification
algorithms. The scatter in the relation between SO and FoF-identified
halos is sure to confuse the interpretation of LSS projections, and
can lead to an overestimation of the latter. Since we were unable to
apply our matched filter algorithm in 3D (we used a $8192^2$ FFT in
2D, and so would require a $8192^3$ FFT in 3D), we avoided the
confusion of SO-FoF by examining the change induced by correlated
projections in density peaks projected in slabs of various
thicknesses.

\par In \S\ref{Results} we measured the projected-peak function in
slabs of 26, 51, 102, 256, 512 $\Mpc$. The use of the compensated
filter rendered the 2D mass function quite stable with the slab
thickness. We then measured the mass evolution of peaks when projected
in the different slabs. We found that the average change in mass of
peaks measured in slabs of 26 $\Mpc$ and 102 $\Mpc$, compared to their
mass in the 51 $\Mpc$ slab, is no greater than 2\%. For the thicker
slabs of 256 $\Mpc$ and 512 $\Mpc$, the average change is also less
than 4\%; we expect these thicker slabs to be affected by uncorrelated
projections. The halo model suggests that a peak measured in a 51
$\Mpc$ slab should suffer a change in mass smaller than 0.1\% compared
to when it is measured in a 26 $\Mpc$ slab. So these changes are
larger than the halo model predictions. On the other hand, we have
tested that the absolute change in the mass of halos is also at the
percent level, in accord with the halo model results. Let us emphasize
that all these findings depend on the choice of filter: our filter
reduces dramatically the impact of correlated projections compared to
other filters. The importance of the filter function in reducing the
LSS contamination of WL cluster mass measurements was first suggested
by \citet{2005A&A...442..851M}. We conclude that with a proper choice
of filter, correlated projections are not a major contaminant for
WL-cluster mass measurements. Uncorrelated projections are likely to
be the dominant contributor to the projection bias.

We also confirm the main result of our earlier study
\citet{2009ApJ...698L..33M}: the projected peak abundance has the same
scaling with cosmology as the 3D and ST mass functions. This is valid
for all the slab thicknesses that we probed. We proposed a very simple
fitting procedure for the 2D mass function, adopting the functional
form of the ST prediction. Following the work of
\citet{2006ApJ...646..881W}, we have estimated the free parameters of
the ST model, by maximizing the extended Poisson likelihood. For all
cosmologies, the fits have an accuracy of up to 10\%-20\% at low and
intermediate mass bins. We also used only the fit for the fiducial
cosmology 2D mass function, to predict the 2D mass functions for the
other cosmologies. In this case, the accuracy was of about 20\% in the
same mass range.

Our goal for the future is to extend our analysis to uncorrelated
projections, and to study both theoretical and numerical avenues for
the prediction of the shear-peak abundance, and its cosmology
dependence. Should such an undertaking be successful, it will
establish the WL method for cluster detection as a very reliable way
to extract cosmological constraints, limited only by survey
systematics.

\acknowledgments 
We thank Peter Schneider for carefully reading the manuscript and for
his suggestions. LM acknowledges the hospitality of the ITP,
University of Z\"{u}rich, where some of this project has been
developed. RES kindly thanks the Argelander Institute, University of
Bonn, for hospitality whilst some of this work was being done. We are
grateful to V. Springel for making public {\tt GADGET-2} and for
providing his {\tt B-FoF} halo finder; and to R. Scoccimarro for
making public his {\tt 2LPT} code. LM is supported by the Deutsche
Forschungsgemeinschaft under the Transregion TRR33 The Dark
Universe. RES acknowledges support from a Marie Curie Reintegration
Grant and the Swiss National Foundation under contract
200021-116696/1. GMB is supported in this work by grant AST-0607667
from the National Science Foundation, Department of Energy grant
DOE-DE-FG02-95ER40893, and NASA grant BEFS-04-0014-0018.



\end{document}